\documentclass[aps,pre,floats,twocolumn,showpacs,superscriptaddress]{revtex4-1}

\usepackage{natbib}

\usepackage{amsmath, amsthm, amssymb}
\usepackage{enumitem}
\usepackage{graphics,graphicx}% Include figure files
\usepackage{epsfig}
\usepackage{times}% Times fonts
\usepackage{dcolumn}% Align table columns on decimal point
\usepackage{url}
\usepackage{color}
\usepackage{bm}

\usepackage[all]{xy}
\usepackage{mdwlist}

\begin{document}

\title{Influence of trust in the spreading of information}

\author{Hongrun Wu}
\affiliation{State Key Laboratory of Software Engineering, Wuhan University, 430072 Wuhan, China}
\affiliation{Departament d'Enginyeria Inform\`atica i Matem\`atiques, Universitat Rovira i Virgili, 43007 Tarragona, Spain}

\author{Alex Arenas}
\affiliation{Departament d'Enginyeria Inform\`atica i Matem\`atiques, Universitat Rovira i Virgili, 43007 Tarragona, Spain}
%\affiliation{IPHES, Institut Catal\`a de Paleoecologia Humana i Evoluci\'o Social, 43007 Tarragona, Spain}

\author{Sergio G\'omez}
\affiliation{Departament d'Enginyeria Inform\`atica i Matem\`atiques, Universitat Rovira i Virgili, 43007 Tarragona, Spain}

\begin{abstract}
The understanding and prediction of information diffusion processes on networks is a major challenge in network theory with many implications in social sciences. Many theoretical advances occurred due to stochastic spreading models. Nevertheless, these stochastic models overlooked the influence of rational decisions on the outcome of the process. For instance, different levels of trust in acquaintances do play a role in information spreading, and actors may change their spreading decisions during the information diffusion process accordingly. Here, we study an information-spreading model in which the decision to transmit or not is based on trust. We explore the interplay between the propagation of information and the trust dynamics happening on a two-layer multiplex network. Actors' trustable or untrustable states are defined as accumulated cooperation or defection behaviors, respectively, in a Prisoner's Dilemma set up, and they are controlled by a memory span. The propagation of information is abstracted as a threshold model on the information-spreading layer, where the threshold depends on the trustability of agents. The analysis of the model is performed using a tree approximation and validated on homogeneous and heterogeneous networks. The results show that the memory of previous actions has a significant effect on the spreading of information. For example, the less memory that is considered, the higher is the diffusion. Information is highly promoted by the emergence of trustable acquaintances. These results provide insight into the effect of plausible biases on spreading dynamics in a multilevel networked system.
\end{abstract}

\pacs{%
89.65.-s,	%Social and economic systems
89.75.Fb,	%Structures and organization in complex systems
89.75.Hc  %Networks and genealogical trees
}

\maketitle

\section{Introduction}

Epidemics, innovation, rumors, gossips and opinions spread on social networks \cite{Daley-1964,Goffman-1964, Goldenberg-2001,Leskovec-2007,Watts-2007,Christakis-2007,Newman-2010,AndreaMontanari-2010,Berger-2012,Pinto-2013,Kreindler-2014,Granell-2013,Lima-2015,Gruhl-2004,Moreno-2004,Kostka-2008,Ostilli-2010,Yang-2010,matamalas2015strategical}. With the availability of large-scale data on social networks, the study of modeling information, rumor and gossip diffusion has recently attracted a great deal of attention \cite{Gruhl-2004,Moreno-2004,Kostka-2008,Ostilli-2010,Yang-2010,Lind-2007,Trpevski-2010,lv-2011,Zinoviev-2011}.

A fundamental step toward understanding information diffusion was the adoption of threshold and cascade models \cite{Schelling-1971-1,Schelling-1971-2,Granovetter-1987} applied to spreading in social networks. In this last context, an actor's diffusion behavior depends on the number of other individuals already engaged in the process \cite{Watts-2002,Singh-2013,Watts-2007}. These works have recently been generalized to multilevel networks \cite{Brummitt12,Lee14,Brummit15,Burkholz16}. The diffusion of information has also been modeled using game theory, giving players a positive payoff if they spread the information \cite{Jackson07,AndreaMontanari-2010,Zinoviev-2010,Zinoviev-2011}. Moreover, the topological factors that may affect the spreading have been investigated in detail in \cite{Pastor-2001,Gleeson-2007,Watts-2002}. Other applications related to information diffusion, such as finding the influential spreaders \cite{Jamali-2009}, maximizing or restraining the information propagation \cite{Kimura-2009,Kempe-2003}, predicting the information diffusion on real-world social networks \cite{Szabo-2010,Pinto-2013,Weng-2013,Cheng-2014}, and revealing general patterns of the diffusion of the temporal information \cite{Yang-2011}, have also been studied extensively.

Nevertheless, most of the existing works on information diffusion focus mainly on network structure, while the decision on whether or not transmit the information in a conscious manner is ignored. However, certainly a person's decision on whether or not to transmit a given piece of information relies on many factors, one of them being trust in the sources \cite{Sherchan-2013}. In a network context, where spreading takes place using the topological structure, this trust is conceived at the level of individual nodes. In our approach, we consider that the decision whether to transmit a certain piece of information depends on the number of trustable and untrustable acquaintances involved in the process. For example, a person who receives a piece of information from a few untrustable neighbors may decide not to forward it because of the lack of reliable evidence, but the behavior changes if these neighbors are trustable. Within this scenario, an untrustable individual is one that you do not know if you can trust; other meanings could be assigned to untrustability, but this one is the more appropriate in our framework.

The interactions of being trustable or untrustable can be modeled using game theory. In this paper, we consider an evolutionary game on graphs \cite{NOWAK-1992,santos2005scale,santos2006evolutionary,GyorgySzab-2007} to model the trustable and untrustable interactions, and the propagation of the information is abstracted as a modified version of the threshold model \cite{Granovetter-1987,Watts-2002} in which trustable neighbors are given more credit in the decision of activation. Similar to \cite{Granell-2013,Granell-2014}, each dynamic takes place in a different layer of a multiplex network, thus taking into account the possibility of having different connections between individuals in the trust and information-spreading layers, respectively. Note that the interaction between both layers is limited to the following: each actor has a game dynamics that specifies the state of being trustable or not. This state is visible for the actors in the spreading phase that takes place on the second layer. In this sense, the trust in one layer influences the spreading on the other.

The paper is structured as follows. In Section~\ref{sect:model} we describe the proposed model and how the information propagation and trust dynamics work. Section~\ref{sect:analysis} describes a tree approximation theory to predict the fraction of active individuals in the population (spreaders) of the proposed model. Section~\ref{sect:results} introduces the validation of the tree approximation theory and the numerical results of the proposed model on random and scale-free multiplex networks. Section~\ref{sect:conclusion} contains some concluding remarks.

\section{\label{sect:model}Trust-Driven Information-Spreading model}

\begin{figure}[!t]
  \begin{center}
  \begin{tabular}{l}
    \mbox{\includegraphics*[width=0.9\columnwidth]{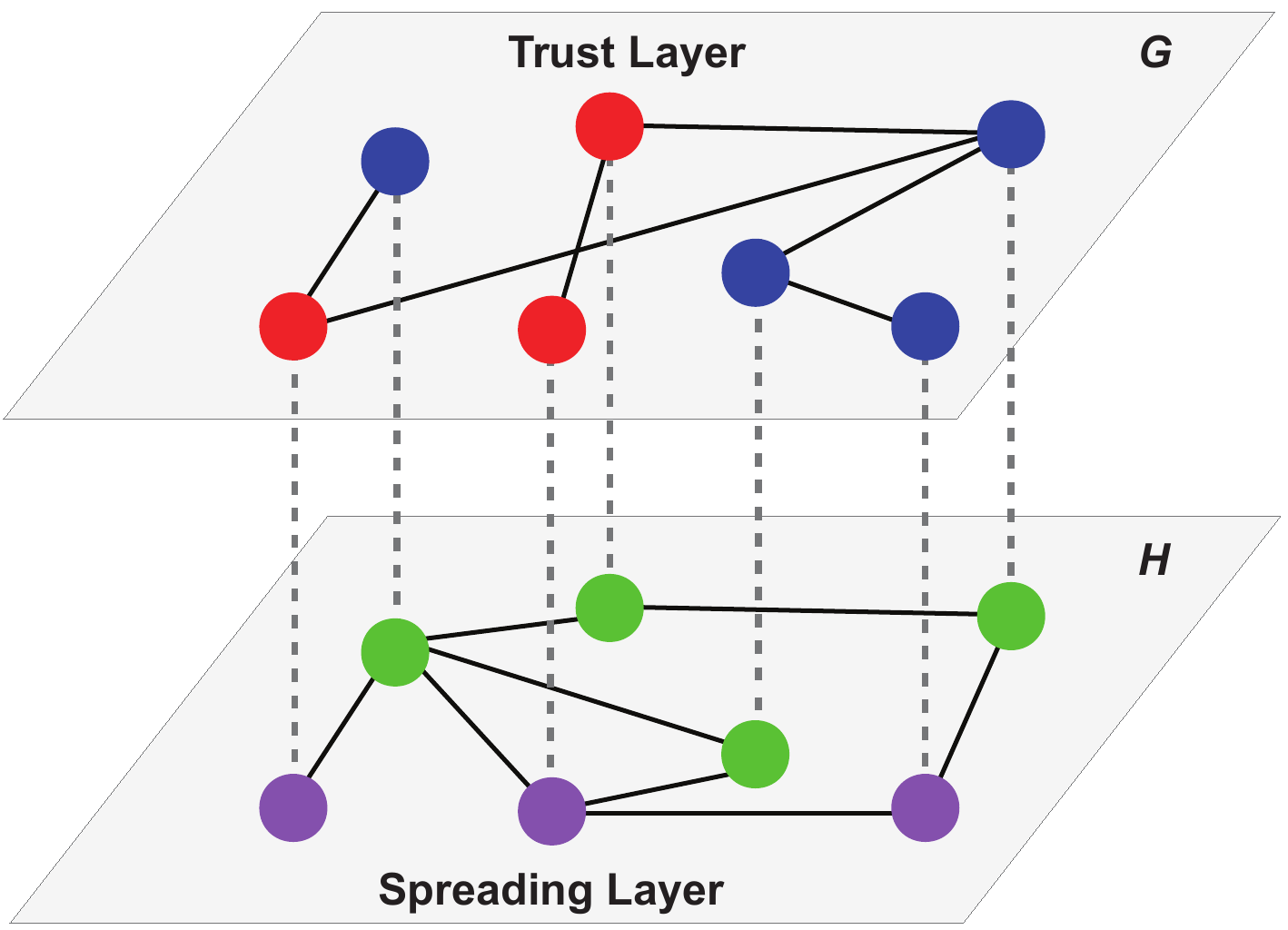}}
  \end{tabular}
  \end{center}
  \caption{Schematic representation of the Trust-Driven Information Spreading (TDIS) model. Each individual is present in both layers of the multiplex (dashed lines connect the representation of the same actor in each layer), and connections with other nodes (solid lines) are different in each layer. Layer $G$ is the trust (game) layer, with nodes in states Trustable (blue) or Untrustable (red) and links represent the acquaintances used in the game, while layer $H$ is the information-spreading layer, with nodes in Active (green) or Inactive (purple) states, and links refer to the possible contacts to whom the information is spread. Every node at the spreading layer uses the information of the states in the trust layer.}
  \label{fig:multiplex_schema}
\end{figure}

We are interested in the introduction of a model of information spreading in which one's individual decision to become an active spreader is influenced by trust in one's neighbors. A common model of choice for information spreading is the threshold model \cite{Granovetter-1987,Watts-2002}, in which an Inactive individual becomes Active whenever the fraction of its active neighbors is above a certain threshold $\theta$. Thus, in order to take into account the trust in the neighbors, we have modified the threshold rule incorporating into it the number of trustable~(T) and untrustable~(U) neighbors, either active or inactive. For the trust evolution we have chosen a Prisoner's Dilemma \cite{Roca-2009,GyorgySzab-2007,Shakarian-2012} evolutionary game dynamics, in which we assimilate persistent cooperators (defectors) as trustable (untrustable) individuals. We show a schematic representation of the system in Fig.~\ref{fig:multiplex_schema}, where $G$ and $H$ represent the game (trust) and information-spreading layers, respectively. The connectivity between individuals is encoded in the corresponding adjacency matrices $G_{ij}$ and $H_{ij}$, which take the value $1$ if nodes~$i$ and~$j$ are connected, and~$0$ otherwise.

This {\em Trust-Driven Information Spreading} (TDIS) model works as an iterative two-stage process on the two-layer multiplex network: one step of the Prisoner's Dilemma game to update the trustable and untrustable states of nodes in layer~$G$, followed by one step of the information spreading in which some of the inactive nodes (non-spreaders) in layer~$H$ become active (spreaders). Note that, according to the semantics of our model, the interaction between layers is one-directional, from the trust to the information-spreading layer, but not in the opposite direction. Other extensions are possible, e.g., bidirectional dependencies, which could produce a coordination between layers \cite{wang2012evolution}, but they are beyond the scope of the present work.

We next introduce the details of the rules governing our TDIS model.

\subsection{\label{subsec:game}Trust dynamics}

Each individual may adopt, in the trust layer, one of two possible strategies, regardless of its state in the information dynamics layer: cooperation~(C), which will be used to define a trustable individual, or defection~(D), for untrustable ones. The benefit of an actor~$i$, playing a weak Prisoner's Dilemma game with one of its neighbors, depends on their game strategies, as dictated by a general payoff matrix \cite{GyorgySzab-2007,NOWAK-1992}:
\begin{equation}
\bordermatrix{
  & C & D \cr
C & 1 & 0 \cr
D & b & 0 \cr}
\label{tab:payoffMatrixWPD}
\end{equation}
where $b \geqslant 1$ is the temptation to defect, i.e., the payoff obtained by a defector when playing with a cooperator.

In every round of the game, each player interacts with all its neighbors on the trust layer, and collects its corresponding accumulated payoff, $\Pi_i$, as the sum of the payoffs of all its game interactions. For the next round, the strategy is updated according to the Replicator Rule \cite{Lieberman09}: player~$i$ imitates the strategy of player~$j$ with probability
\begin{equation} \label{eq:replicator}
P_{i \rightarrow j}= \frac{\Pi_{j}-\Pi_{i}}{b\,\max(k_{i},k_{j})}\,,
\end{equation}
where $k_{i}$ is the degree (number of neighbors) of node~$i$. After all players have had the opportunity to update their strategies, the players' payoffs are reset to zero and a new round of the game starts.

Since trust is usually a long term opinion on people, we do not directly describe trustable individuals as cooperators (and untrustable as defectors), but rather as the most common behavior in a certain time span $\Delta t \geqslant 1$. More precisely, an individual is considered Trustable (T) if, in the last $\Delta t$ time steps, it has been acting as a cooperator more than 50\% of the times, and Untrustable (U) otherwise. Only when $\Delta t = 1$ do the concepts of cooperation and trust coincide. The choice of 50\% is rather flexible, and changing this value simply unbalances the system towards more cooperation or defection, nevertheless we think this choice is a plausible bound.

\subsection{\label{subsec:info}Information spreading dynamics}

Once the trustability of all nodes has been established in the previous stage of the TDIS model, one step of the information-spreading dynamics takes place. Each Inactive individual becomes Active whenever the {\em influence} $I(\bm{m},\bm{k};\alpha)$ from its neighbors is larger than a threshold $\theta$ ($0\leqslant \theta \leqslant 1$), where
\begin{equation}
  I(\bm{m},\bm{k};\alpha) \equiv
  \frac{(1+\alpha) m_{T}+(1-\alpha)m_{U}}{(1+\alpha) k_T+(1-\alpha) k_U}\,.
  \label{eq:influence}
\end{equation}
Here, $k_T$ ($k_U$) is the number of trustable (untrustable) neighbors, and $m_T$ ($m_U$) is the number of active trustable (untrustable) neighbors, satisfying $m_T \leqslant k_T$ and $m_U \leqslant k_U$. For convenience, we have defined the activity and degree vectors, $\bm{m}=(m_{T},m_{U})$ and $\bm{k}=(k_{T},k_{U})$, respectively. Influence parameter $\alpha$ ($0\leqslant \alpha \leqslant 1$) controls the level of influence of trustable neighbors versus untrustable ones, i.e., the larger $\alpha$ is, the more importance we give to our trustable neighbors in the decision between becoming active or remaining inactive. There are two important particular cases:
\begin{itemize}
\item When $\alpha=0$, the standard threshold model is recovered \cite{Watts-2002},
  \begin{equation}
    I(\bm{m},\bm{k};\alpha=0) = \frac{m_T+m_U}{k_T+k_U} = \frac{m}{k}\,,
  \end{equation}
  where $k=k_T+k_U$ is the total degree of the node, and $m=m_T+m_U$ is the number of active neighbors. In this case, the trust layer decouples from the information-spreading layer, thus having two independent dynamics.
\item When $\alpha=1$, only trustable neighbors are taken into account,
  \begin{equation}
    I(\bm{m},\bm{k};\alpha=1) = \frac{m_T}{k_T}\,.
  \end{equation}
\end{itemize}
For intermediate values of the influence parameter, both trustable and untrustable neighbors, with different levels of influence, contribute to the activation condition $I(\bm{m},\bm{k};\alpha) > \theta$.
Note that the spreading process is still deterministic but influenced via threshold by the trustability state of nodes. In this sense, the spreader mimics the decision-making process to spread according to a certain threshold based on trustability.

\section{\label{sect:analysis}Analysis of the information spreading}

The information-spreading process in our model belongs to the class of binary-state dynamics, the transmitting rate of which can be described as depending on the number of nearest neighbors in the two possible active or inactive states. In this section, we make use of the tree approximation theory \cite{Gleeson-2008,Melnik-2011,Yagan-2012} to predict the extension of the information spreading when trust in the neighbors is taken into account. We first analyze the case in which the distribution of trustable and untrustable nodes does not change, i.e., when there is no trust dynamics.

In the tree approximation approach, the network is supposed to have a locally treelike structure, with low clustering. A random node is selected as the root of the tree, using the degree distribution of the network $p(\bm{k})$, and the rest of the tree is built by following edges satisfying the joint degree distribution $P(\bm{k},\bm{k}')$, i.e.\ the probability that a randomly chosen edge connects two nodes of degrees $\bm{k}$ and $\bm{k}'$. The cascade of information is started by an initial fraction $\rho_{0}$ of active nodes, distributed by degree as $\rho_{0}^{(\bm{k})}$, thus satisfying $\rho_{0}=\sum_{\bm{k}}\rho_{0}^{(\bm{k})}$. The analysis starts in the leaves of the tree (level $n=0$), and goes up towards the root, which is reached at level $n\to\infty$. Alternatively and equivalently, we could consider $n$ as a time step in a synchronous update of the spreading dynamics through the tree; here we prefer to use the tree-level interpretation to avoid confusion with the time in the TDIS dynamics.

The expected fraction of active nodes $\rho$ at the steady state can be obtained by considering the root:
\begin{equation}
  \rho = \sum_{\bm{k}} p(\bm{k}) \rho^{(\bm{k})}\,,
  \label{eq:rho}
\end{equation}
where
\begin{equation}
  \rho^{(\bm{k})} = \rho_{0}^{(\bm{k})} + (1-\rho_{0}^{(\bm{k})})
    \sum_{m_{T}=0}^{k_{T}}\sum_{m_{U}=0}^{k_{U}} F(\bm{m},\bm{k}) B_{\infty}(\bm{m},\bm{k})\,.
  \label{eq:rhok}
\end{equation}
This equation expresses that the root with degree $\bm{k}$ is active either if it was initially active (with probability $\rho_{0}^{(\bm{k})}$), or if it was initially inactive (with probability $1-\rho_{0}^{(\bm{k})}$) but was activated by its children. The probability of activation from the children has two terms: the probability $B_{\infty}(\bm{m},\bm{k})$ of having $\bm{m}$ active neighbors among the total $\bm{k}$ children (i.e., $m_{T}$ active trustable neighbors and $m_{U}$ active untrustable neighbors, among the total $k_{T}$ trustable and $k_{U}$ untrustable children), and the probability $F(\bm{m},\bm{k})$ that these $\bm{m}$ children activate the parent. For our information-spreading dynamics, the response function $F(\bm{m},\bm{k})$ is just
\begin{equation}
  F(\bm{m},\bm{k})=
  \begin{cases}
    1 & \text{ if } I(\bm{m},\bm{k};\alpha) > \theta\,, \\
    0 & \text{ otherwise, }
  \end{cases}
  \label{eq:fFun}
\end{equation}
and the distribution of active children can be written as the product of two independent binomial distributions, one for the trustable children and the other for the untrustable ones:
\begin{align}
  B_{\infty}(\bm{m},\bm{k}) =
    &\binom{k_{T}}{m_{T}} \, (r_\infty^{(\bm{k},T)})^{m_{T}} \, (1-r_\infty^{(\bm{k},T)})^{k_{T}-m_{T}} \nonumber
    \\
    \times
    &\binom{k_{U}}{m_{U}} \, (r_\infty^{(\bm{k},U)})^{m_{U}} \, (1-r_\infty^{(\bm{k},U)})^{k_{U}-m_{U}}\,.
  \label{eq:fBInf}
\end{align}
Variable $r_n^{(\bm{k},T)}$ ($r_n^{(\bm{k},U)}$) represents the probability that a trustable (untrustable) child of an inactive level $n$ node of degree $k$ is active. We can put them in terms of the probabilities $q_n^{(\bm{k},T)}$ ($q_n^{(\bm{k},U)}$) that a level $n$ trustable (untrustable) node of degree $\bm{k}$ is active conditional on its parent being inactive, leading to
\begin{equation}
  r_n^{(\bm{k},T)} = \frac{\sum_{\bm{k}'} P(\bm{k},\bm{k}')\,q_n^{(\bm{k}',T)}}
                          {\sum_{\bm{k}'} P(\bm{k},\bm{k}')}\,,
  \label{eq:rtkC}
\end{equation}
\begin{equation}
  r_n^{(\bm{k},U)} = \frac{\sum_{\bm{k}'} P(\bm{k},\bm{k}')\,q_n^{(\bm{k}',U)}}
                          {\sum_{\bm{k}'} P(\bm{k},\bm{k}')}\,.
  \label{eq:rtkD}
\end{equation}

Probabilities $q_{n+1}^{(\bm{k},T)}$ and $q_{n+1}^{(\bm{k},U)}$ of a node on level $n+1$ satisfy expressions similar to Eq.~(\ref{eq:rhok}):
\begin{align}
  q_{n+1}^{(\bm{k},T)} = &\rho_{0}^{(\bm{k})} + (1-\rho_{0}^{(\bm{k})}) \nonumber
    \\
    & \times \left[
      \frac{k_{T}\, p(\bm{k},T)}{z^{(T)}}
      \sum_{m_{T}=0}^{k_{T}-1} \sum_{m_{U}=0}^{k_{U}} F(\bm{m},\bm{k}) B_n^{(T)}(\bm{m},\bm{k})
      \right. \nonumber
    \\
    & \left.
      {} +
      \frac{k_{U}\, p(\bm{k},T)}{z^{(T)}}
      \sum_{m_{T}=0}^{k_{T}} \sum_{m_{U}=0}^{k_{U}-1} F(\bm{m},\bm{k}) B_n^{(U)}(\bm{m},\bm{k})
      \right],
  \label{eq:qtkc}
\end{align}
\begin{align}
  q_{n+1}^{(\bm{k},U)} = &\rho_{0}^{(\bm{k})} + (1-\rho_{0}^{(\bm{k})}) \nonumber
    \\
    & \times \left[
      \frac{k_{T}\, p(\bm{k},U)}{z^{(U)}}
      \sum_{m_{T}=0}^{k_{T}-1} \sum_{m_{U}=0}^{k_{U}} F(\bm{m},\bm{k}) B_n^{(T)}(\bm{m},\bm{k})
      \right. \nonumber
    \\
    & \left.
      {} +
      \frac{k_{U}\, p(\bm{k},U)}{z^{(U)}}
      \sum_{m_{T}=0}^{k_{T}} \sum_{m_{U}=0}^{k_{U}-1} F(\bm{m},\bm{k}) B_n^{(U)}(\bm{m},\bm{k})
      \right],
  \label{eq:qtkD}
\end{align}
where $z^{(T)} = \sum_{\bm{k}} k\,p(\bm{k},T)$ and $z^{(U)} = \sum_{\bm{k}} k\,p(\bm{k},U)$ are the average degrees of trustable and untrustable nodes, respectively, and
\begin{align}
  B_n^{(T)}(\bm{m},\bm{k}) =
    &\binom{k_{T}-1}{m_{T}} \, (r_n^{(\bm{k},T)})^{m_{T}} \, (1-r_n^{(\bm{k},T)})^{k_{T}-m_{T}-1} \nonumber
    \\
    \times
    &\binom{k_{U}}{m_{U}} \, (r_n^{(\bm{k},U)})^{m_{U}} \, (1-r_n^{(\bm{k},U)})^{k_{U}-m_{U}}\,,
  \label{eq:fBnt}
\end{align}
\begin{align}
  B_n^{(U)}(\bm{m},\bm{k}) =
    &\binom{k_{T}}{m_{T}} \, (r_n^{(\bm{k},T)})^{m_{T}} \, (1-r_n^{(\bm{k},T)})^{k_{T}-m_{T}} \nonumber
    \\
    \times
    &\binom{k_{U}-1}{m_{U}} \, (r_n^{(\bm{k},U)})^{m_{U}} \, (1-r_n^{(\bm{k},U)})^{k_{U}-m_{U}-1}.
  \label{eq:fBnu}
\end{align}
The idea is that, for a node at level $n+1$ with degree $\bm{k}$, we have to substract the parent from the list of inactive neighbors, due to the definitions of $q_{n+1}^{(\bm{k},T)}$ and $q_{n+1}^{(\bm{k},U)}$. Additionally, the degree distribution in the tree is, by construction, equivalent to the degree distribution of the nearest neighbors in the original network, and it also depends on whether the node is either trustable or untrustable. For instance, the probability of having a trustable node in the tree with degree $\bm{k}$ is equal to $k\, p(\bm{k},T)/z_T$, which reduces to $k_{T}\, p(\bm{k},T)/z_T$ when you add the condition that the parent is trustable, and $k_{U}\, p(\bm{k},T)/z_T$ for an untrustable parent.

The solution of these equations is obtained by just starting with an initial condition for $q_{0}^{(\bm{k},T)}$ and $q_{0}^{(\bm{k},U)}$, and iterating them ($n=1,2,3,\ldots$) until a stationary value is reached. If we suppose the fraction of trustable nodes $s^{(T)}$ is fixed and uncorrelated with the initial fraction of active nodes $\rho_{0}^{(\bm{k})}$, then we can set $q_{0}^{(\bm{k},T)} = \rho_{0}^{(\bm{k})}\,s^{(T)}$ and $q_{0}^{(\bm{k},U)} = \rho_{0}^{(\bm{k})}\,(1-s^{(T)})$. Note that the two main conditions for the applicability of the tree approximation approach are satisfied, namely the permanently active property (i.e., active nodes cannot become inactive) and the nondecreasing character of the response function $F(\bm{m},\bm{k})$ for increasing values of $\bm{m}$ (maintaining $\bm{k}$ fixed); see \cite{Gleeson-2007,Gleeson-2008}.

In our full TDIS model, the trust dynamics is independent of the information spreading, but the opposite does not hold, i.e.\ the number and distribution of trustable nodes may affect the size of the information cascade. Moreover, the trust dynamics makes the distribution of trustable individuals change in time, thus interfering with the information spreading. However, since we suppose the trust is a long-term effect of the game dynamics, governed by the memory span $\Delta t$, we may use an adiabatic approximation in which, for each time step $t$ of the trust dynamics, we calculate the extension $\rho(t)$ of the information spreading as if it were instantaneous (using as input to the equations the current distribution of trustable and untrustable nodes). The final outcome is then $\rho = \max_t \rho(t)$, meaning that, during the trust evolution, there is a moment at which the information spreading extends the largest, and the rest of the TDIS dynamics is not able to increase it any more. Unfortunately, we do not know in advance when this peak of spreading is going to happen, and simulations show that in many occasions it appears in the transitory of the trust evolution, thus forcing us to follow the whole trust dynamics to perform the prediction of the information spreading.

\section{\label{sect:results}Results}

We first validate the accuracy of the tree approximation theory in Sect.~\ref{subsec:valid}, then we show the results for the TDIS model in Sect.~\ref{subsec:tdisresults}, and finally we explore the effect of the distribution of trustable and untrustable nodes on the information spreading in Sect.~\ref{subsec:randomize}. Throughout the simulations, we make use of two reference multiplex networks, both with $N=1000$ nodes and average degree $z=6$ in each layer. In the first multiplex the layers are random Erd\H{o}s-R\'{e}nyi networks (ER), and in the second they are scale-free networks built using the Barab{\'a}si-Albert model (BA); in both cases the trust and information-spreading layers are uncorrelated between them. We also set the threshold value to $\theta=0.3$, the initial fraction of spreaders in the information layer to $1\%$ ($\rho_{0}=0.01$), and we make 150~repetitions of the Monte Carlo simulations, each one consisting of 20000~time steps. Note that a classical threshold dynamics for these networks and with the selected threshold has no global cascades; see \cite{Watts-2002}.

\subsection{\label{subsec:valid}Validation of the tree approximation}

\begin{figure}[!t]
  \begin{center}
  \begin{tabular}{l}
     \mbox{\includegraphics*[width=\columnwidth]{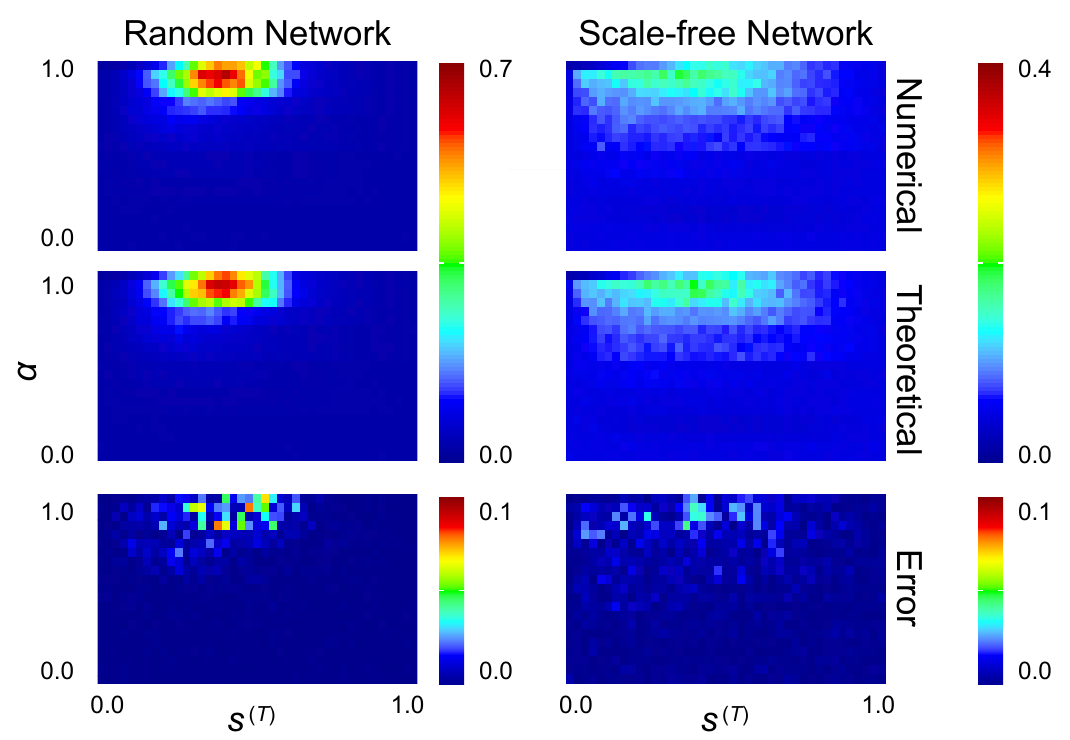}}
  \end{tabular}
  \end{center}
  \caption{Comparison of the fraction of active nodes $\rho$ (in color code) obtained with Monte Carlo simulations (first row) and with theoretical predictions using the tree approximation (second row), for a fixed and randomly distributed fraction $s^{(T)}$ of trustable nodes, and varying values of the influence parameter $\alpha$. The left column corresponds to a random ER network, and the right corresponds to a scale-free BA network. The third row shows the absolute difference between theory and Monte Carlo results, which amounts to a global relative error of $6.4\%$ (ER) and $8.3\%$ (BA), respectively. }
  \label{fig:valid}
\end{figure}

In Fig.~\ref{fig:valid} we compare the fraction of active nodes $\rho$ predicted using the tree approximation in Sect.~\ref{sect:analysis} with respect to Monte Carlo simulations. A fraction $s^{(T)}$ of trustable nodes is randomly assigned, for varying values of $s^{(T)}$ and of the influence parameter $\alpha$ in the range $[0,1]$. The state of the nodes, trustable or untrustable, remains fixed during the information spreading dynamics. The results show a good agreement between the tree approximation predictions and the Monte Carlo simulations.

Note that, for the current set-up, if the influence of trustable and untrustable nodes is the same (i.e.\ $\alpha=0$, which recovers the standard threshold model), the information spreading is very low, almost negligible. However, as $\alpha$ increases, the trust in the neighbors allows larger spreading, with maximums of $\rho$ at about $0.7$ (ER) and $0.2$ (BA), respectively, for certain values of the influence parameter and of the fraction of trustable nodes. The differences between the ER and BA networks are just a higher level of diffusion for the ER network. Therefore, we may state that the trust in the neighbors enhances the information spreading, provided the fraction of trustable nodes is neither too large nor too small. The same behavior is also observed in the next section, when the full TDIS is taken into account.

\subsection{\label{subsec:tdisresults}Results for the trust-driven information spreading model}

\begin{figure}[!t]
  \begin{center}
  \begin{tabular}{l}
    (a) \\
    \mbox{\includegraphics*[width=.95\columnwidth]{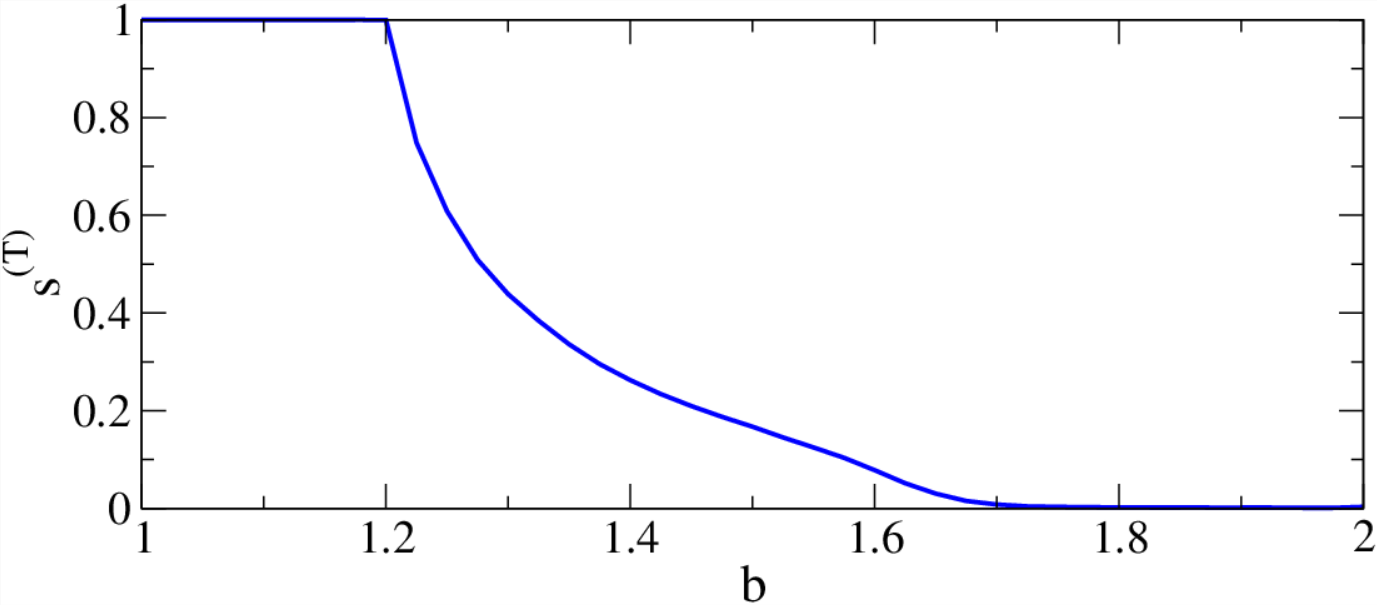}} \\
    (b) \\
    \mbox{\includegraphics*[width=.95\columnwidth]{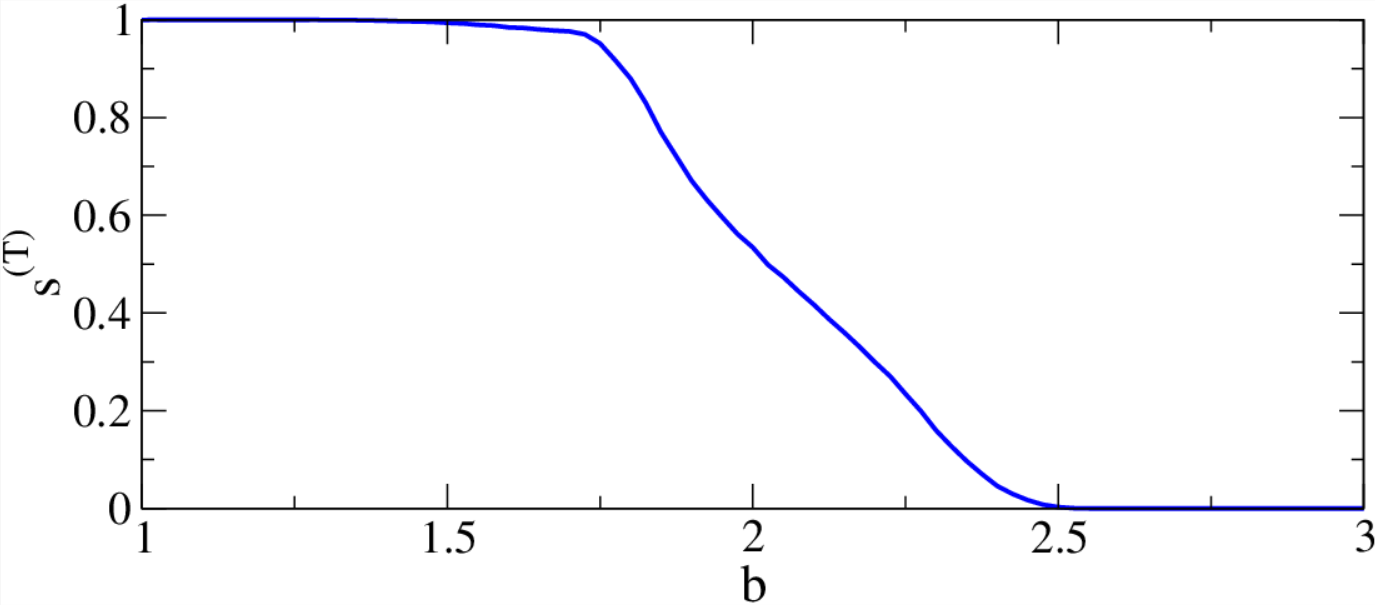}}
  \end{tabular}
  \end{center}
  \caption{Fraction of trustable nodes as a function of the temptation for the two considered networks with average degree $z=6$: (a) random ER network; (b) scale-free BA network. The parameters of the Monte Carlo simulations are as follows: memory span $\Delta t = 1$, 50\% of initial randomly distributed trustable nodes, and average of the fraction of trustable nodes over all the repetitions and for the last 1000~time steps.}
  \label{fig:trust}
\end{figure}

As explained in Sect.~\ref{sect:model}, the TDIS model has a trust dynamics based on a weak Prisoner's Dilemma, which depends on the temptation parameter $b$ and the memory time span $\Delta t$. In Fig.~\ref{fig:trust} we show as a reference the average fraction of trustable nodes in the stationary state for the two considered networks, with a memory span $\Delta t = 1$ and an initial fraction of cooperators at $50\%$. It shows that, tuning the temptation, we are able to scan from full trustability $s^{(T)}=1$ to full untrustability $s^{(T)}=0$. For larger values of $\Delta t$, the fraction of trustable nodes does not change in a significant way. Note that we have selected $b\in[1,2]$ for the ER network and $b\in[1,3]$ for the BA network.

\begin{figure*}[!t]
  \begin{center}
  \begin{tabular}{l}
    (a) \\
    \mbox{\includegraphics*[width=.80\textwidth]{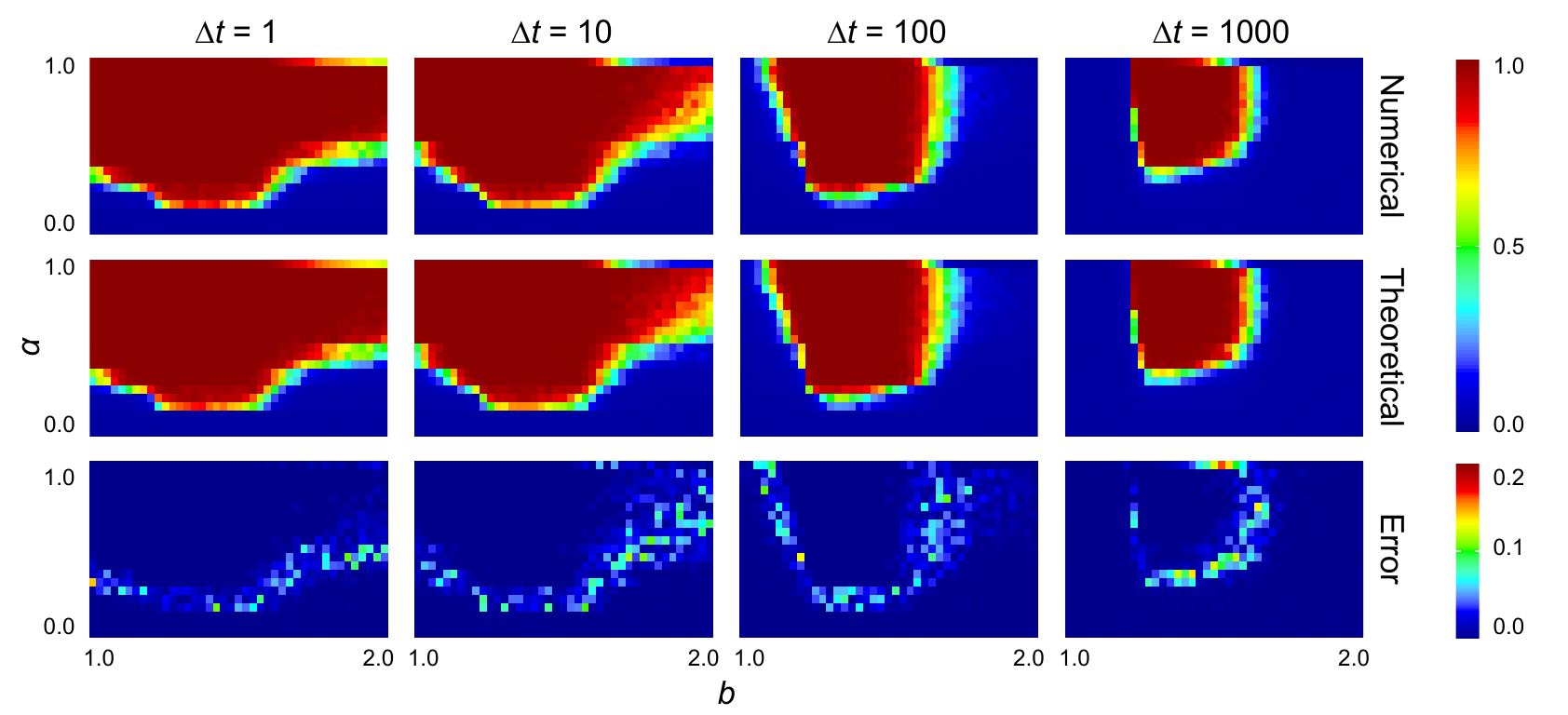}} \\
    (b) \\
    \mbox{\includegraphics*[width=.80\textwidth]{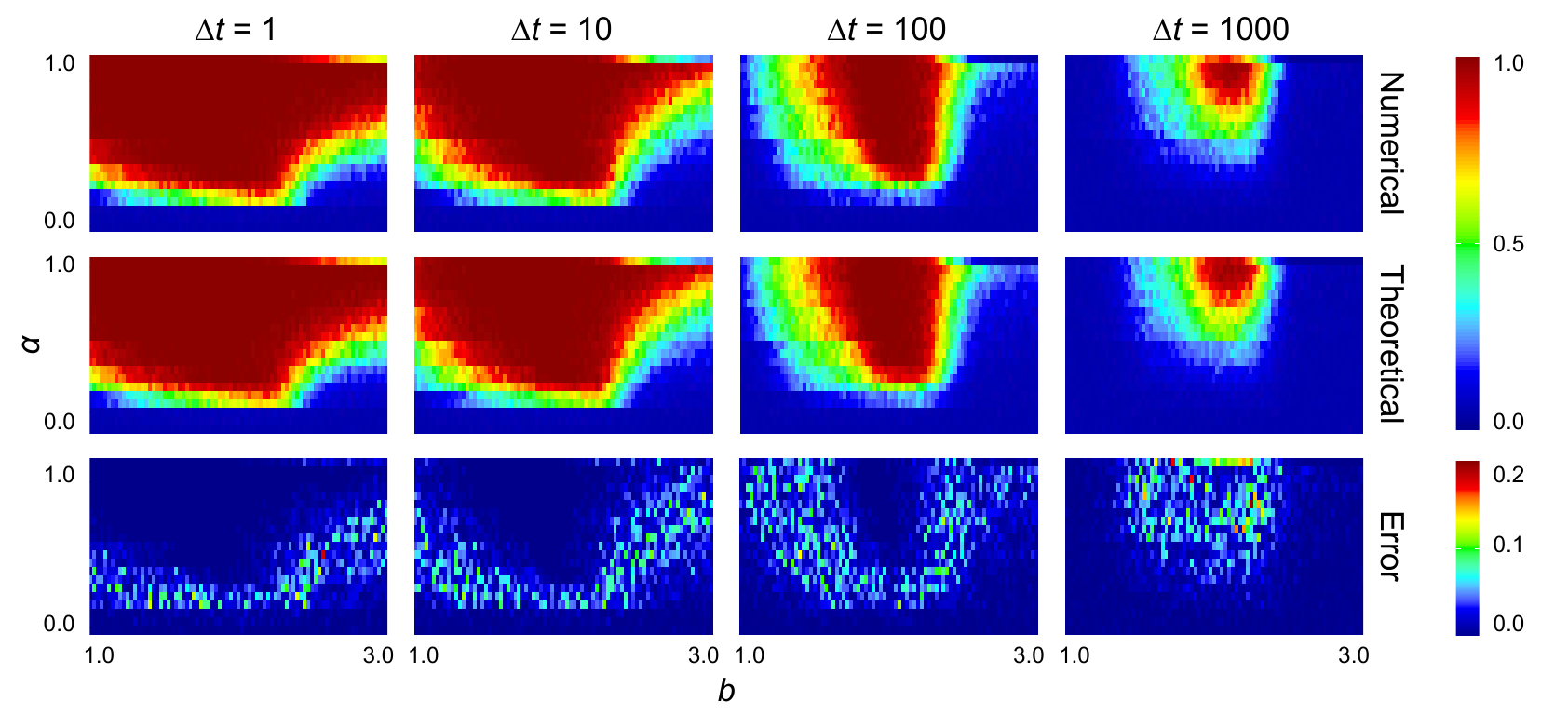}}
  \end{tabular}
  \end{center}
  \caption{Fraction of active nodes $\rho$ (in color code) as a function of the temptation $b$ and the influence $\alpha$ for the full TDIS model running on top of the two considered multiplex networks: (a) random ER multiplex; (b) scale-free BA multiplex. In each panel, the first row corresponds to Monte Carlo simulations, the second row to the theoretical predictions using the tree approximation, and the last one to the absolute difference between them. The columns correspond to four different values of the memory span $\Delta t$, with respective global relative errors between theory and Monte Carlo of 1.6\%, 1.8\%, 2.4\% and 5.1\% for the ER multiplex, and 3.3\%, 3.3\%, 4.2\% and 9.3\% for the BA multiplex. Same Monte Carlo parameters as in Fig.~\ref{fig:trust}, and the fraction of active nodes is the average of their values in the last time step over all the repetitions.}
  \label{fig:rho}
\end{figure*}

Figure~\ref{fig:rho} shows a comparison between the Monte Carlo and theoretical predictions of the fraction of active nodes for the two considered multiplex networks (ER and BA), for different values of the temptation $b$, the influence $\alpha$, and the memory span $\Delta t$ parameters. The theoretical prediction is performed using the scheme specified at the end of Sect.~\ref{sect:analysis}. The agreement between them is remarkable, with global relative errors ranging from 1.6\% in the best case (ER, $\Delta t = 1$) to 9.3\% in the worst one (BA, $\Delta t = 1000$).

We first observe in Fig.~\ref{fig:rho} that, for the current threshold $\theta=0.3$, the activity does not spread when there is no distinction between trustable and untrustable nodes ($\alpha=0$). However, when we start increasing the influence of trustable neighbors, the fraction of active nodes quickly rises, easily covering the whole population. This effect is more important for intermediate values of the temptation $b$ and shorter memory time spans. Taking into account the results in Figs.~\ref{fig:valid} and~\ref{fig:trust}, it becomes evident that the information spreading is not just a consequence of the fraction of trustable nodes in the stationary state of the trust dynamics, but it must also depend on its transitory states. For example, when $b$ is close to 1, all nodes are trustable in the stady state ($s^{(T)}=1$), which corresponds to a region in Fig.~\ref{fig:valid} where the activity cannot propagate, but it does. Therefore, the only possibility is that, during the transitory of the trust dynamics, the number of trustable nodes changes continuously from the initial fraction 50\% to a final value (not necessarily in a monotonic way), and it is in these intermediate states when the information spreading achieves its maximum level. Moreover, the fact that the influence $\alpha$ can be small for full spreading unlike in Fig.~\ref{fig:valid}, points to the importance of the whole trust dynamics to explain the spreading, which cannot be understood by just accounting for the fraction of trustable individuals.

The reduction of information spreading as memory grows can also be explained through its effect on the transitory of the trust dynamics: from the point of view of the information spreading dynamics, when $\Delta t$ grows, the fluctuations of the transitory are smoothed, and consequently nodes remain in the same state (trustable or untrustable) for longer times. Thus, the fraction of trustable nodes approaches a constant value that, in the limit $\Delta t \to \infty$, is equal to the stationary value in Fig.~\ref{fig:trust}. In this situation, we expect to have a level of spreading similar to those in Fig.~\ref{fig:valid}, hence explaining the important reduction in information spreading as memory rises.

The results are equivalent for both the ER and BA multiplex networks, except for a larger and smoother transition region (in the parameters space) between no spreading and full diffusion for the BA multiplex, whereas the ER presents a much abrupt boundary between them.

In summary, we have seen that, in simple terms, trust in the neighbors helps in the spreading of information, while long term memory in the assignment of trust restrains it. Moreover, major spreading is accomplished at intermediate values of the temptation, for which the population of trustable and untrustable individuals is more balanced.

\subsection{\label{subsec:randomize}The effect of the distribution of trustable individuals}

\begin{figure*}[!t]
  \begin{center}
  \begin{tabular}{l}
    (a) \\
    \mbox{\includegraphics*[width=.80\textwidth]{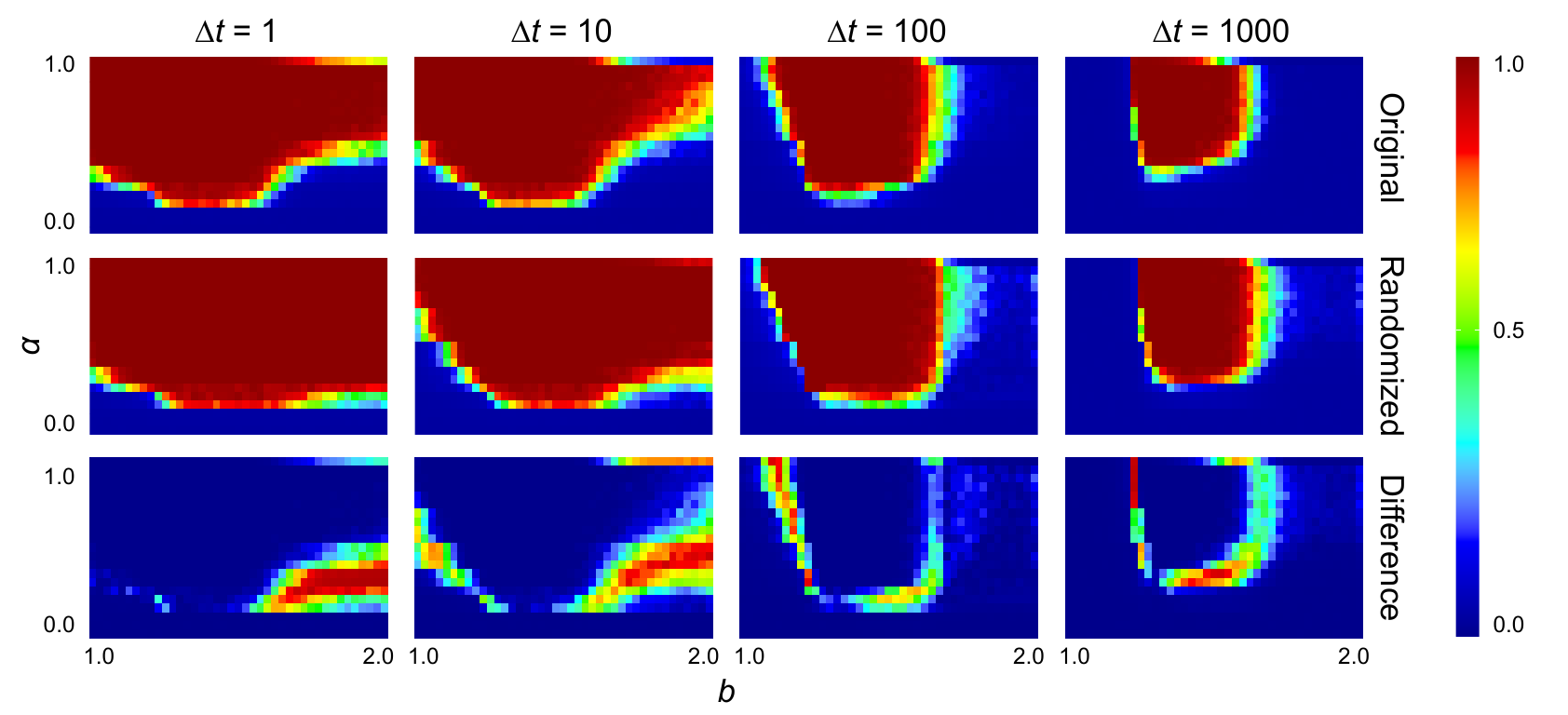}} \\
    (b) \\
    \mbox{\includegraphics*[width=.80\textwidth]{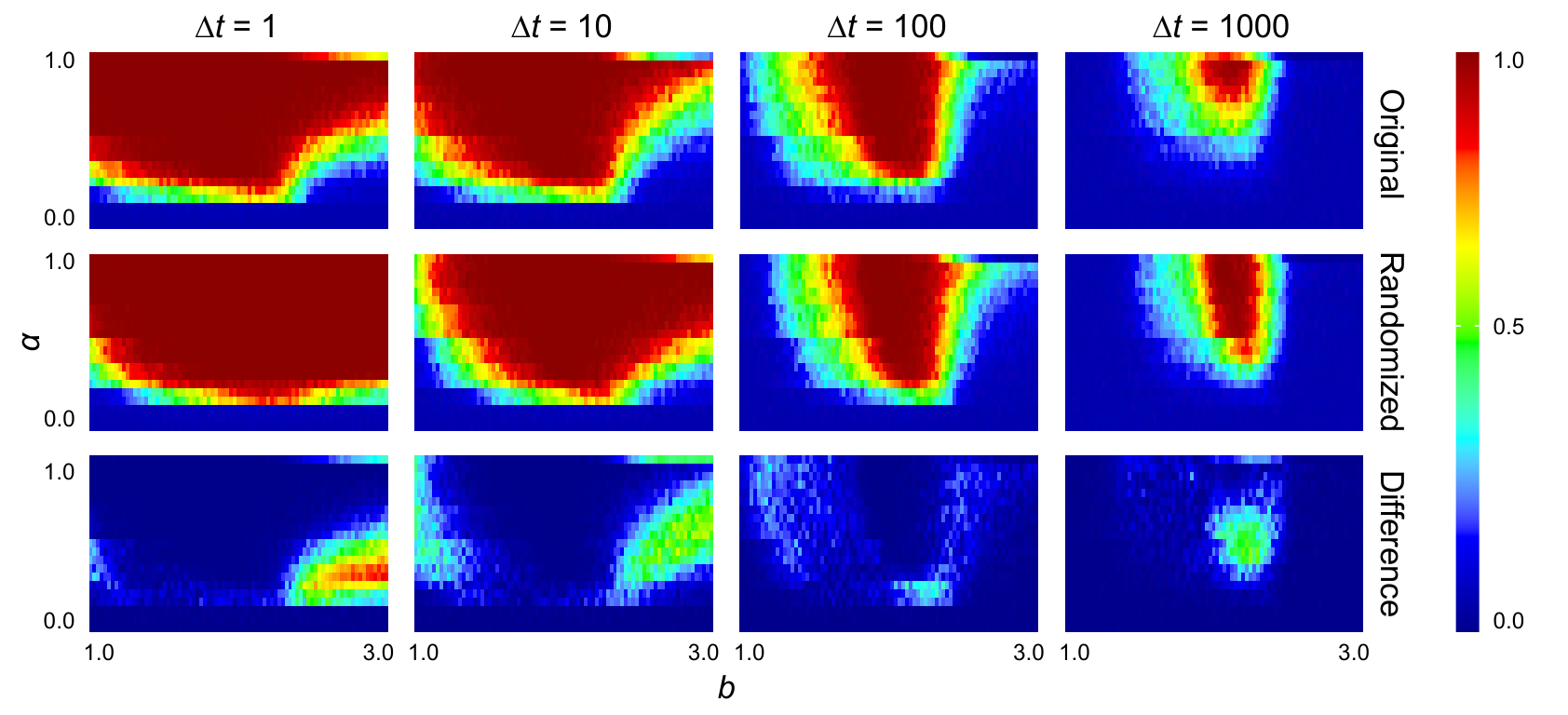}}
  \end{tabular}
  \end{center}
  \caption{Dependence of the fraction of active nodes $\rho$ (in color code) on the distribution of trustable individuals, for the two considered multiplex networks: (a) a random ER multiplex; (b) a scale-free BA multiplex. In each panel, the first row corresponds to Monte Carlo simulations of the TDIS model, the second row corresponds to Monte Carlo simulations in which the assignment of trustable nodes is randomized, and the last one corresponds to the absolute difference between them. The columns correspond to four different values of the memory span $\Delta t$.}
  \label{fig:randomized}
\end{figure*}

Although the fraction of trustable individuals is very important for the spreading of information, as established in the previous Sects.~\ref{subsec:valid} and~\ref{subsec:tdisresults}, it remains to be seen if their distribution across the network produced by the trust dynamics is also relevant or not.
To this end, we compare Monte Carlo simulations of the full TDIS model (first rows in Fig.~\ref{fig:randomized}) to new simulations with randomized assignments of the trust states (second rows in Fig.~\ref{fig:randomized}). More precisely, for every TDIS Monte Carlo simulation, we build a randomized instance in which, for each time step, the fraction of trustable nodes is preserved but their distribution is randomly reshuffled.

The results in Fig.~\ref{fig:randomized} show a general enlargement of the regions with maximum information spreading, thus confirming the importance of the distribution of the trustable nodes. Since the fraction of trustable nodes is not enough to account for the final spreading of the information, we cannot avoid the tracking of the trust dynamics to obtain good predictions, using the procedure in Sect.~\ref{sect:analysis}. Again, both the ER and BA multiplex networks show a qualitatively similar behavior.

%The global relative difference between them is 35\%, 36\%, 28\% and 52\% for the ER multiplex, and 26\%, 22\%, 16\% and 30\% for the BA multiplex.

\section{\label{sect:conclusion}Conclusion}

In this work, we have introduced a model of information spreading, based on the standard threshold model, which takes into account trust in the neighbors in the decision on whether or not to spread the information. Three factors affecting this personal behavior are taken into account: the degree of influence of trustable acquaintances versus untrustable ones, the memory span to consider trustable individuals as such, and the temptation to not cooperate in the next action. The result is a trust-driven information spreading model, on top of a two-layer multiplex network, in which individuals participate in two processes, a trust dynamics in one layer and an information spreading in the other. The influence and distribution of trustable actors affect the diffusion of information, while there is no interaction in the opposite direction, from the spreading to the trust layer. For the diffusion of the information, individuals become spreaders when the influence of their neighbors exceeds a certain threshold. We have shown that this model allows an analytical treatment of the information diffusion, based on a tree approximation, in good agreement with Monte Carlo simulations.

The results show how an increasing influence of trustable neighbors promotes information diffusion, which is easily diffused to all the population. However, the information is restrained when a long term memory of previous behaviors is used to assign the trustable or untrustable character of individuals. Additionally, intermediate values of the temptation enhance the spreading thanks to the fact that they yield balanced populations of trustable and untrustable individuals, which is the most favorable configuration to satisfy the threshold condition and become a spreader. We have also shown that not only the fraction but also the distribution and evolution of trustable nodes are important to predict the final outcome of the spreading process, and that all the previous results apply both for random ER and heterogeneous scale-free BA multiplex networks. These results provide clues to understand and quantify the effects of a rational individual's decision making in the propagation of all kinds of information.

\section{Acknowledgements}
We thank J.\ Poncela and J.\ Matamalas for their help. H.W. acknowledges the financial support of the National Basic Research Program of China (grant number 2014CB340404). A.A. and S.G. acknowledge MULTIPLEX (grant number 317532) of the European Commission, and the Spanish Ministerio de Econom\'{\i}a y Competitividad (grant number FIS2015-71582-C2-1). A.A. also acknowledges ICREA Academia and the James S. McDonnell Foundation.

%\bibliography{references}

\begin{thebibliography}{56}%
\makeatletter
\providecommand \@ifxundefined [1]{%
 \@ifx{#1\undefined}
}%
\providecommand \@ifnum [1]{%
 \ifnum #1\expandafter \@firstoftwo
 \else \expandafter \@secondoftwo
 \fi
}%
\providecommand \@ifx [1]{%
 \ifx #1\expandafter \@firstoftwo
 \else \expandafter \@secondoftwo
 \fi
}%
\providecommand \natexlab [1]{#1}%
\providecommand \enquote  [1]{``#1''}%
\providecommand \bibnamefont  [1]{#1}%
\providecommand \bibfnamefont [1]{#1}%
\providecommand \citenamefont [1]{#1}%
\providecommand \href@noop [0]{\@secondoftwo}%
\providecommand \href [0]{\begingroup \@sanitize@url \@href}%
\providecommand \@href[1]{\@@startlink{#1}\@@href}%
\providecommand \@@href[1]{\endgroup#1\@@endlink}%
\providecommand \@sanitize@url [0]{\catcode `\\12\catcode `\$12\catcode
  `\&12\catcode `\#12\catcode `\^12\catcode `\_12\catcode `\%12\relax}%
\providecommand \@@startlink[1]{}%
\providecommand \@@endlink[0]{}%
\providecommand \url  [0]{\begingroup\@sanitize@url \@url }%
\providecommand \@url [1]{\endgroup\@href {#1}{\urlprefix }}%
\providecommand \urlprefix  [0]{URL }%
\providecommand \Eprint [0]{\href }%
\providecommand \doibase [0]{http://dx.doi.org/}%
\providecommand \selectlanguage [0]{\@gobble}%
\providecommand \bibinfo  [0]{\@secondoftwo}%
\providecommand \bibfield  [0]{\@secondoftwo}%
\providecommand \translation [1]{[#1]}%
\providecommand \BibitemOpen [0]{}%
\providecommand \bibitemStop [0]{}%
\providecommand \bibitemNoStop [0]{.\EOS\space}%
\providecommand \EOS [0]{\spacefactor3000\relax}%
\providecommand \BibitemShut  [1]{\csname bibitem#1\endcsname}%
\let\auto@bib@innerbib\@empty
%</preamble>
\bibitem [{\citenamefont {Daley}\ and\ \citenamefont
  {Kendall}(1964)}]{Daley-1964}%
  \BibitemOpen
  \bibfield  {author} {\bibinfo {author} {\bibfnamefont {D.~J.}\ \bibnamefont
  {Daley}}\ and\ \bibinfo {author} {\bibfnamefont {D.~G.}\ \bibnamefont
  {Kendall}},\ }\href {\doibase 10.1038/2041118a0} {\bibfield  {journal}
  {\bibinfo  {journal} {Nature}\ }\textbf {\bibinfo {volume} {204}},\ \bibinfo
  {pages} {1118} (\bibinfo {year} {1964})}\BibitemShut {NoStop}%
\bibitem [{\citenamefont {Goffman}\ and\ \citenamefont
  {Newill}(1964)}]{Goffman-1964}%
  \BibitemOpen
  \bibfield  {author} {\bibinfo {author} {\bibfnamefont {W.}~\bibnamefont
  {Goffman}}\ and\ \bibinfo {author} {\bibfnamefont {V.~A.}\ \bibnamefont
  {Newill}},\ }\href {\doibase 10.1038/204225a0} {\bibfield  {journal}
  {\bibinfo  {journal} {Nature}\ }\textbf {\bibinfo {volume} {204}},\ \bibinfo
  {pages} {225} (\bibinfo {year} {1964})}\BibitemShut {NoStop}%
\bibitem [{\citenamefont {Goldenberg}\ \emph {et~al.}(2001)\citenamefont
  {Goldenberg}, \citenamefont {Libai},\ and\ \citenamefont
  {Muller}}]{Goldenberg-2001}%
  \BibitemOpen
  \bibfield  {author} {\bibinfo {author} {\bibfnamefont {J.}~\bibnamefont
  {Goldenberg}}, \bibinfo {author} {\bibfnamefont {B.}~\bibnamefont {Libai}}, \
  and\ \bibinfo {author} {\bibfnamefont {E.}~\bibnamefont {Muller}},\ }\href
  {\doibase 10.1023/A:1011122126881} {\bibfield  {journal} {\bibinfo  {journal}
  {Marketing Lett.}\ }\textbf {\bibinfo {volume} {12}},\ \bibinfo {pages} {211}
  (\bibinfo {year} {2001})}\BibitemShut {NoStop}%
\bibitem [{\citenamefont {Leskovec}\ \emph {et~al.}(2007)\citenamefont
  {Leskovec}, \citenamefont {Adamic},\ and\ \citenamefont
  {Huberman}}]{Leskovec-2007}%
  \BibitemOpen
  \bibfield  {author} {\bibinfo {author} {\bibfnamefont {J.}~\bibnamefont
  {Leskovec}}, \bibinfo {author} {\bibfnamefont {L.~A.}\ \bibnamefont
  {Adamic}}, \ and\ \bibinfo {author} {\bibfnamefont {B.~A.}\ \bibnamefont
  {Huberman}},\ }\href@noop {} {\bibfield  {journal} {\bibinfo  {journal} {ACM
  Trans. Web}\ }\textbf {\bibinfo {volume} {1}},\ \bibinfo {pages} {5}
  (\bibinfo {year} {2007})}\BibitemShut {NoStop}%
\bibitem [{\citenamefont {Watts}\ and\ \citenamefont
  {Dodds}(2007)}]{Watts-2007}%
  \BibitemOpen
  \bibfield  {author} {\bibinfo {author} {\bibfnamefont {D.~J.}\ \bibnamefont
  {Watts}}\ and\ \bibinfo {author} {\bibfnamefont {P.~S.}\ \bibnamefont
  {Dodds}},\ }\href {\doibase 10.1086/518527} {\bibfield  {journal} {\bibinfo
  {journal} {J. Consum. Res.}\ }\textbf {\bibinfo {volume} {374}},\ \bibinfo
  {pages} {441} (\bibinfo {year} {2007})}\BibitemShut {NoStop}%
\bibitem [{\citenamefont {Christakis}\ and\ \citenamefont
  {Fowler}(2007)}]{Christakis-2007}%
  \BibitemOpen
  \bibfield  {author} {\bibinfo {author} {\bibfnamefont {N.~A.}\ \bibnamefont
  {Christakis}}\ and\ \bibinfo {author} {\bibfnamefont {J.~H.}\ \bibnamefont
  {Fowler}},\ }\href {\doibase 10.1056/NEJMsa066082} {\bibfield  {journal}
  {\bibinfo  {journal} {New Engl. J. Med.}\ }\textbf {\bibinfo {volume}
  {357}},\ \bibinfo {pages} {370} (\bibinfo {year} {2007})}\BibitemShut
  {NoStop}%
\bibitem [{\citenamefont {Newman}(2010)}]{Newman-2010}%
  \BibitemOpen
  \bibfield  {author} {\bibinfo {author} {\bibfnamefont {M.}~\bibnamefont
  {Newman}},\ }\href@noop {} {\emph {\bibinfo {title} {Networks: An
  Introduction}}}\ (\bibinfo  {publisher} {Oxford University Press},\ \bibinfo
  {year} {2010})\BibitemShut {NoStop}%
\bibitem [{\citenamefont {Montanari}\ and\ \citenamefont
  {Saberi}(2010)}]{AndreaMontanari-2010}%
  \BibitemOpen
  \bibfield  {author} {\bibinfo {author} {\bibfnamefont {A.}~\bibnamefont
  {Montanari}}\ and\ \bibinfo {author} {\bibfnamefont {A.}~\bibnamefont
  {Saberi}},\ }\href@noop {} {\bibfield  {journal} {\bibinfo  {journal} {Proc.
  Natl. Acad. Sci. USA.}\ }\textbf {\bibinfo {volume} {107}},\ \bibinfo {pages}
  {20196} (\bibinfo {year} {2010})}\BibitemShut {NoStop}%
\bibitem [{\citenamefont {Berger}\ and\ \citenamefont
  {Milkman}(2012)}]{Berger-2012}%
  \BibitemOpen
  \bibfield  {author} {\bibinfo {author} {\bibfnamefont {J.}~\bibnamefont
  {Berger}}\ and\ \bibinfo {author} {\bibfnamefont {K.~L.}\ \bibnamefont
  {Milkman}},\ }\href {\doibase 10.1509/jmr.10.0353} {\bibfield  {journal}
  {\bibinfo  {journal} {J. Marketing Res.}\ }\textbf {\bibinfo {volume} {49}},\
  \bibinfo {pages} {192} (\bibinfo {year} {2012})}\BibitemShut {NoStop}%
\bibitem [{\citenamefont {Pinto}\ \emph {et~al.}(2013)\citenamefont {Pinto},
  \citenamefont {Almeida},\ and\ \citenamefont {Gon\'alves}}]{Pinto-2013}%
  \BibitemOpen
  \bibfield  {author} {\bibinfo {author} {\bibfnamefont {H.}~\bibnamefont
  {Pinto}}, \bibinfo {author} {\bibfnamefont {J.~M.}\ \bibnamefont {Almeida}},
  \ and\ \bibinfo {author} {\bibfnamefont {M.~A.}\ \bibnamefont {Gon\'alves}},\
  }in\ \href@noop {} {\emph {\bibinfo {booktitle} {Proceedings of the sixth ACM
  International Conference on Web Search and Data Mining}}}\ (\bibinfo
  {address} {Rome, Italy},\ \bibinfo {year} {2013})\ pp.\ \bibinfo {pages}
  {365--374}\BibitemShut {NoStop}%
\bibitem [{\citenamefont {Kreindler}\ and\ \citenamefont
  {Young}(2014)}]{Kreindler-2014}%
  \BibitemOpen
  \bibfield  {author} {\bibinfo {author} {\bibfnamefont {G.~E.}\ \bibnamefont
  {Kreindler}}\ and\ \bibinfo {author} {\bibfnamefont {H.~P.}\ \bibnamefont
  {Young}},\ }\href {\doibase 10.1073/pnas.1400842111} {\bibfield  {journal}
  {\bibinfo  {journal} {Proc. Natl. Acad. Sci. USA}\ }\textbf {\bibinfo
  {volume} {111}},\ \bibinfo {pages} {10881} (\bibinfo {year}
  {2014})}\BibitemShut {NoStop}%
\bibitem [{\citenamefont {Granell}\ \emph {et~al.}(2013)\citenamefont
  {Granell}, \citenamefont {G\'omez},\ and\ \citenamefont
  {Arenas}}]{Granell-2013}%
  \BibitemOpen
  \bibfield  {author} {\bibinfo {author} {\bibfnamefont {C.}~\bibnamefont
  {Granell}}, \bibinfo {author} {\bibfnamefont {S.}~\bibnamefont {G\'omez}}, \
  and\ \bibinfo {author} {\bibfnamefont {A.}~\bibnamefont {Arenas}},\ }\href
  {\doibase 10.1103/PhysRevLett.111.128701} {\bibfield  {journal} {\bibinfo
  {journal} {Phys. Rev. Lett.}\ }\textbf {\bibinfo {volume} {111}},\ \bibinfo
  {pages} {128701} (\bibinfo {year} {2013})}\BibitemShut {NoStop}%
\bibitem [{\citenamefont {Lima}\ \emph {et~al.}(2015)\citenamefont {Lima},
  \citenamefont {De~Domenico}, \citenamefont {Pejovic},\ and\ \citenamefont
  {Musolesi}}]{Lima-2015}%
  \BibitemOpen
  \bibfield  {author} {\bibinfo {author} {\bibfnamefont {A.}~\bibnamefont
  {Lima}}, \bibinfo {author} {\bibfnamefont {M.}~\bibnamefont {De~Domenico}},
  \bibinfo {author} {\bibfnamefont {V.}~\bibnamefont {Pejovic}}, \ and\
  \bibinfo {author} {\bibfnamefont {M.}~\bibnamefont {Musolesi}},\ }\href
  {\doibase 10.1038/srep10650} {\bibfield  {journal} {\bibinfo  {journal} {Sci.
  Rep.}\ }\textbf {\bibinfo {volume} {5}},\ \bibinfo {pages} {10650} (\bibinfo
  {year} {2015})}\BibitemShut {NoStop}%
\bibitem [{\citenamefont {Gruhl}\ \emph {et~al.}(2004)\citenamefont {Gruhl},
  \citenamefont {Guha}, \citenamefont {Liben-Nowell},\ and\ \citenamefont
  {Tomkins}}]{Gruhl-2004}%
  \BibitemOpen
  \bibfield  {author} {\bibinfo {author} {\bibfnamefont {D.}~\bibnamefont
  {Gruhl}}, \bibinfo {author} {\bibfnamefont {R.}~\bibnamefont {Guha}},
  \bibinfo {author} {\bibfnamefont {D.}~\bibnamefont {Liben-Nowell}}, \ and\
  \bibinfo {author} {\bibfnamefont {A.}~\bibnamefont {Tomkins}},\ }in\
  \href@noop {} {\emph {\bibinfo {booktitle} {Proceedings of the 13th
  International Conference on World Wide Web}}}\ (\bibinfo {address} {New York,
  USA},\ \bibinfo {year} {2004})\ pp.\ \bibinfo {pages} {491--501}\BibitemShut
  {NoStop}%
\bibitem [{\citenamefont {Moreno}\ \emph {et~al.}(2004)\citenamefont {Moreno},
  \citenamefont {Nekovee},\ and\ \citenamefont {Pacheco}}]{Moreno-2004}%
  \BibitemOpen
  \bibfield  {author} {\bibinfo {author} {\bibfnamefont {Y.}~\bibnamefont
  {Moreno}}, \bibinfo {author} {\bibfnamefont {M.}~\bibnamefont {Nekovee}}, \
  and\ \bibinfo {author} {\bibfnamefont {A.~F.}\ \bibnamefont {Pacheco}},\
  }\href {\doibase 10.1103/PhysRevE.69.066130} {\bibfield  {journal} {\bibinfo
  {journal} {Phys. Rev. E}\ }\textbf {\bibinfo {volume} {69}},\ \bibinfo
  {pages} {066130} (\bibinfo {year} {2004})}\BibitemShut {NoStop}%
\bibitem [{\citenamefont {Kostka}\ \emph {et~al.}(2008)\citenamefont {Kostka},
  \citenamefont {Oswald},\ and\ \citenamefont {Wattenhofer}}]{Kostka-2008}%
  \BibitemOpen
  \bibfield  {author} {\bibinfo {author} {\bibfnamefont {J.}~\bibnamefont
  {Kostka}}, \bibinfo {author} {\bibfnamefont {Y.~A.}\ \bibnamefont {Oswald}},
  \ and\ \bibinfo {author} {\bibfnamefont {R.}~\bibnamefont {Wattenhofer}},\
  }in\ \href@noop {} {\emph {\bibinfo {booktitle} {Structural Information and
  Communication Complexity}}}\ (\bibinfo  {publisher} {Springer},\ \bibinfo
  {year} {2008})\ pp.\ \bibinfo {pages} {185--196}\BibitemShut {NoStop}%
\bibitem [{\citenamefont {Ostilli}\ \emph {et~al.}(2010)\citenamefont
  {Ostilli}, \citenamefont {Yoneki}, \citenamefont {Leung}, \citenamefont
  {Mendes}, \citenamefont {Pietro},\ and\ \citenamefont
  {Crowcroft}}]{Ostilli-2010}%
  \BibitemOpen
  \bibfield  {author} {\bibinfo {author} {\bibfnamefont {M.}~\bibnamefont
  {Ostilli}}, \bibinfo {author} {\bibfnamefont {E.}~\bibnamefont {Yoneki}},
  \bibinfo {author} {\bibfnamefont {I.~X.}\ \bibnamefont {Leung}}, \bibinfo
  {author} {\bibfnamefont {J.~F.}\ \bibnamefont {Mendes}}, \bibinfo {author}
  {\bibfnamefont {L.}~\bibnamefont {Pietro}}, \ and\ \bibinfo {author}
  {\bibfnamefont {J.}~\bibnamefont {Crowcroft}},\ }\href {\doibase
  10.1016/j.procs.2010.04.262} {\bibfield  {journal} {\bibinfo  {journal}
  {Procedia Comput. Sci.}\ }\textbf {\bibinfo {volume} {1}},\ \bibinfo {pages}
  {2331} (\bibinfo {year} {2010})}\BibitemShut {NoStop}%
\bibitem [{\citenamefont {Yang}\ and\ \citenamefont
  {Leskovec}(2010)}]{Yang-2010}%
  \BibitemOpen
  \bibfield  {author} {\bibinfo {author} {\bibfnamefont {J.}~\bibnamefont
  {Yang}}\ and\ \bibinfo {author} {\bibfnamefont {J.}~\bibnamefont
  {Leskovec}},\ }in\ \href@noop {} {\emph {\bibinfo {booktitle} {Proceedings of
  the 2010 IEEE International Conference on Data Mining}}}\ (\bibinfo {address}
  {Sydney, Australia},\ \bibinfo {year} {2010})\ pp.\ \bibinfo {pages}
  {599--608}\BibitemShut {NoStop}%
\bibitem [{\citenamefont {Matamalas}\ \emph {et~al.}(2015)\citenamefont
  {Matamalas}, \citenamefont {Poncela-Casasnovas}, \citenamefont {G{\'o}mez},\
  and\ \citenamefont {Arenas}}]{matamalas2015strategical}%
  \BibitemOpen
  \bibfield  {author} {\bibinfo {author} {\bibfnamefont {J.~T.}\ \bibnamefont
  {Matamalas}}, \bibinfo {author} {\bibfnamefont {J.}~\bibnamefont
  {Poncela-Casasnovas}}, \bibinfo {author} {\bibfnamefont {S.}~\bibnamefont
  {G{\'o}mez}}, \ and\ \bibinfo {author} {\bibfnamefont {A.}~\bibnamefont
  {Arenas}},\ }\href@noop {} {\bibfield  {journal} {\bibinfo  {journal} {Sci.
  Rep.}\ }\textbf {\bibinfo {volume} {5}},\ \bibinfo {pages} {9519} (\bibinfo
  {year} {2015})}\BibitemShut {NoStop}%
\bibitem [{\citenamefont {Lind}\ \emph {et~al.}(2007)\citenamefont {Lind},
  \citenamefont {da~Silva}, \citenamefont {Andrade},\ and\ \citenamefont
  {Herrmann}}]{Lind-2007}%
  \BibitemOpen
  \bibfield  {author} {\bibinfo {author} {\bibfnamefont {P.~G.}\ \bibnamefont
  {Lind}}, \bibinfo {author} {\bibfnamefont {L.~R.}\ \bibnamefont {da~Silva}},
  \bibinfo {author} {\bibfnamefont {J.}~\bibnamefont {Andrade}, \bibfnamefont
  {J.~S.}}, \ and\ \bibinfo {author} {\bibfnamefont {H.~J.}\ \bibnamefont
  {Herrmann}},\ }\href {\doibase 10.1103/PhysRevE.76.036117} {\bibfield
  {journal} {\bibinfo  {journal} {Phys. Rev. E}\ }\textbf {\bibinfo {volume}
  {76}},\ \bibinfo {pages} {036117} (\bibinfo {year} {2007})}\BibitemShut
  {NoStop}%
\bibitem [{\citenamefont {Trpevski}\ \emph {et~al.}(2010)\citenamefont
  {Trpevski}, \citenamefont {Tang},\ and\ \citenamefont
  {Kocarev}}]{Trpevski-2010}%
  \BibitemOpen
  \bibfield  {author} {\bibinfo {author} {\bibfnamefont {D.}~\bibnamefont
  {Trpevski}}, \bibinfo {author} {\bibfnamefont {W.~K.}\ \bibnamefont {Tang}},
  \ and\ \bibinfo {author} {\bibfnamefont {L.}~\bibnamefont {Kocarev}},\ }\href
  {\doibase 10.1103/PhysRevE.81.056102} {\bibfield  {journal} {\bibinfo
  {journal} {Phys. Rev. E}\ }\textbf {\bibinfo {volume} {81}},\ \bibinfo
  {pages} {056102} (\bibinfo {year} {2010})}\BibitemShut {NoStop}%
\bibitem [{\citenamefont {L\"u}\ \emph {et~al.}(2011)\citenamefont {L\"u},
  \citenamefont {Chen},\ and\ \citenamefont {Zhou}}]{lv-2011}%
  \BibitemOpen
  \bibfield  {author} {\bibinfo {author} {\bibfnamefont {L.}~\bibnamefont
  {L\"u}}, \bibinfo {author} {\bibfnamefont {D.-B.}\ \bibnamefont {Chen}}, \
  and\ \bibinfo {author} {\bibfnamefont {T.}~\bibnamefont {Zhou}},\ }\href
  {\doibase 10.1088/1367-2630/13/12/123005} {\bibfield  {journal} {\bibinfo
  {journal} {New J. Phys.}\ }\textbf {\bibinfo {volume} {13}},\ \bibinfo
  {pages} {123005} (\bibinfo {year} {2011})}\BibitemShut {NoStop}%
\bibitem [{\citenamefont {Zinoviev}\ and\ \citenamefont
  {Duong}(2011)}]{Zinoviev-2011}%
  \BibitemOpen
  \bibfield  {author} {\bibinfo {author} {\bibfnamefont {D.}~\bibnamefont
  {Zinoviev}}\ and\ \bibinfo {author} {\bibfnamefont {V.}~\bibnamefont
  {Duong}},\ }in\ \href@noop {} {\emph {\bibinfo {booktitle} {Proceedings of
  the 44th Annual Simulation Symposium}}}\ (\bibinfo {address} {Boston,
  Massachusetts},\ \bibinfo {year} {2011})\ pp.\ \bibinfo {pages}
  {47--52}\BibitemShut {NoStop}%
\bibitem [{\citenamefont {Braun}(1995)}]{Schelling-1971-1}%
  \BibitemOpen
  \bibfield  {author} {\bibinfo {author} {\bibfnamefont {N.}~\bibnamefont
  {Braun}},\ }\href {\doibase 10.1177/1043463195007002005} {\bibfield
  {journal} {\bibinfo  {journal} {Ration. Soc.}\ }\textbf {\bibinfo {volume}
  {7}},\ \bibinfo {pages} {167} (\bibinfo {year} {1995})}\BibitemShut {NoStop}%
\bibitem [{\citenamefont {Granovetter}\ and\ \citenamefont
  {Soong}(1986)}]{Schelling-1971-2}%
  \BibitemOpen
  \bibfield  {author} {\bibinfo {author} {\bibfnamefont {M.}~\bibnamefont
  {Granovetter}}\ and\ \bibinfo {author} {\bibfnamefont {R.}~\bibnamefont
  {Soong}},\ }\href {\doibase 10.1177/1043463195007002005} {\bibfield
  {journal} {\bibinfo  {journal} {J. Econ. Behav. Organ.}\ }\textbf {\bibinfo
  {volume} {7}},\ \bibinfo {pages} {83} (\bibinfo {year} {1986})}\BibitemShut
  {NoStop}%
\bibitem [{\citenamefont {Granovetter}(1987)}]{Granovetter-1987}%
  \BibitemOpen
  \bibfield  {author} {\bibinfo {author} {\bibfnamefont {M.}~\bibnamefont
  {Granovetter}},\ }\href {\doibase 10.1086/226707} {\bibfield  {journal}
  {\bibinfo  {journal} {Am. J. Sociol.}\ }\textbf {\bibinfo {volume} {83}},\
  \bibinfo {pages} {1420} (\bibinfo {year} {1987})}\BibitemShut {NoStop}%
\bibitem [{\citenamefont {Watts}(2002)}]{Watts-2002}%
  \BibitemOpen
  \bibfield  {author} {\bibinfo {author} {\bibfnamefont {D.~J.}\ \bibnamefont
  {Watts}},\ }\href {\doibase 10.1073/pnas.082090499} {\bibfield  {journal}
  {\bibinfo  {journal} {Proc. Natl. Acad. Sci. USA}\ }\textbf {\bibinfo
  {volume} {99}},\ \bibinfo {pages} {5766} (\bibinfo {year}
  {2002})}\BibitemShut {NoStop}%
\bibitem [{\citenamefont {Singh}\ \emph {et~al.}(2013)\citenamefont {Singh},
  \citenamefont {Sreenivasan}, \citenamefont {Szymanski},\ and\ \citenamefont
  {Korniss}}]{Singh-2013}%
  \BibitemOpen
  \bibfield  {author} {\bibinfo {author} {\bibfnamefont {P.}~\bibnamefont
  {Singh}}, \bibinfo {author} {\bibfnamefont {S.}~\bibnamefont {Sreenivasan}},
  \bibinfo {author} {\bibfnamefont {B.~K.}\ \bibnamefont {Szymanski}}, \ and\
  \bibinfo {author} {\bibfnamefont {G.}~\bibnamefont {Korniss}},\ }\href
  {\doibase 10.1038/srep02330} {\bibfield  {journal} {\bibinfo  {journal} {Sci.
  Rep.}\ }\textbf {\bibinfo {volume} {3}},\ \bibinfo {pages} {2330} (\bibinfo
  {year} {2013})}\BibitemShut {NoStop}%
\bibitem [{\citenamefont {Brummitt}\ \emph {et~al.}(2012)\citenamefont
  {Brummitt}, \citenamefont {Lee},\ and\ \citenamefont {Goh}}]{Brummitt12}%
  \BibitemOpen
  \bibfield  {author} {\bibinfo {author} {\bibfnamefont {C.~D.}\ \bibnamefont
  {Brummitt}}, \bibinfo {author} {\bibfnamefont {K.-M.}\ \bibnamefont {Lee}}, \
  and\ \bibinfo {author} {\bibfnamefont {K.-I.}\ \bibnamefont {Goh}},\ }\href
  {\doibase 10.1103/PhysRevE.85.045102} {\bibfield  {journal} {\bibinfo
  {journal} {Phys. Rev. E}\ }\textbf {\bibinfo {volume} {85}},\ \bibinfo
  {pages} {045102} (\bibinfo {year} {2012})}\BibitemShut {NoStop}%
\bibitem [{\citenamefont {Lee}\ \emph {et~al.}(2014)\citenamefont {Lee},
  \citenamefont {Brummitt},\ and\ \citenamefont {Goh}}]{Lee14}%
  \BibitemOpen
  \bibfield  {author} {\bibinfo {author} {\bibfnamefont {K.-M.}\ \bibnamefont
  {Lee}}, \bibinfo {author} {\bibfnamefont {C.~D.}\ \bibnamefont {Brummitt}}, \
  and\ \bibinfo {author} {\bibfnamefont {K.-I.}\ \bibnamefont {Goh}},\ }\href
  {\doibase 10.1103/PhysRevE.90.062816} {\bibfield  {journal} {\bibinfo
  {journal} {Phys. Rev. E}\ }\textbf {\bibinfo {volume} {90}},\ \bibinfo
  {pages} {062816} (\bibinfo {year} {2014})}\BibitemShut {NoStop}%
\bibitem [{\citenamefont {Brummitt}\ and\ \citenamefont
  {Kobayashi}(2015)}]{Brummit15}%
  \BibitemOpen
  \bibfield  {author} {\bibinfo {author} {\bibfnamefont {C.~D.}\ \bibnamefont
  {Brummitt}}\ and\ \bibinfo {author} {\bibfnamefont {T.}~\bibnamefont
  {Kobayashi}},\ }\href {\doibase 10.1103/PhysRevE.91.062813} {\bibfield
  {journal} {\bibinfo  {journal} {Phys. Rev. E}\ }\textbf {\bibinfo {volume}
  {91}},\ \bibinfo {pages} {062813} (\bibinfo {year} {2015})}\BibitemShut
  {NoStop}%
\bibitem [{\citenamefont {Burkholz}\ \emph {et~al.}(2016)\citenamefont
  {Burkholz}, \citenamefont {Leduc}, \citenamefont {Garas},\ and\ \citenamefont
  {Schweitzer}}]{Burkholz16}%
  \BibitemOpen
  \bibfield  {author} {\bibinfo {author} {\bibfnamefont {R.}~\bibnamefont
  {Burkholz}}, \bibinfo {author} {\bibfnamefont {M.~V.}\ \bibnamefont {Leduc}},
  \bibinfo {author} {\bibfnamefont {A.}~\bibnamefont {Garas}}, \ and\ \bibinfo
  {author} {\bibfnamefont {F.}~\bibnamefont {Schweitzer}},\ }\href {\doibase
  http://dx.doi.org/10.1016/j.physd.2015.10.004} {\bibfield  {journal}
  {\bibinfo  {journal} {Physica D}\ }\textbf {\bibinfo {volume} {323--324}},\
  \bibinfo {pages} {64 } (\bibinfo {year} {2016})}\BibitemShut {NoStop}%
\bibitem [{\citenamefont {Jackson}\ and\ \citenamefont
  {Yariv}(2007)}]{Jackson07}%
  \BibitemOpen
  \bibfield  {author} {\bibinfo {author} {\bibfnamefont {M.~O.}\ \bibnamefont
  {Jackson}}\ and\ \bibinfo {author} {\bibfnamefont {L.}~\bibnamefont
  {Yariv}},\ }\href {\doibase 10.1257/aer.97.2.92} {\bibfield  {journal}
  {\bibinfo  {journal} {Am. Econ. Rev.}\ }\textbf {\bibinfo {volume} {97}},\
  \bibinfo {pages} {92} (\bibinfo {year} {2007})}\BibitemShut {NoStop}%
\bibitem [{\citenamefont {Zinoviev}\ and\ \citenamefont
  {Duong}(2010)}]{Zinoviev-2010}%
  \BibitemOpen
  \bibfield  {author} {\bibinfo {author} {\bibfnamefont {D.}~\bibnamefont
  {Zinoviev}}\ and\ \bibinfo {author} {\bibfnamefont {V.}~\bibnamefont
  {Duong}},\ }in\ \href@noop {} {\emph {\bibinfo {booktitle} {Proceedings of
  the 2010 Summer Computer Simulation Conference}}}\ (\bibinfo {address}
  {Ottawa, Canada},\ \bibinfo {year} {2010})\ pp.\ \bibinfo {pages}
  {358--363}\BibitemShut {NoStop}%
\bibitem [{\citenamefont {Pastor-Satorras}\ and\ \citenamefont
  {Vespignani}(2001)}]{Pastor-2001}%
  \BibitemOpen
  \bibfield  {author} {\bibinfo {author} {\bibfnamefont {R.}~\bibnamefont
  {Pastor-Satorras}}\ and\ \bibinfo {author} {\bibfnamefont {A.}~\bibnamefont
  {Vespignani}},\ }\href {\doibase 10.1103/PhysRevLett.86.3200} {\bibfield
  {journal} {\bibinfo  {journal} {Phys. Rev. Lett.}\ }\textbf {\bibinfo
  {volume} {86}},\ \bibinfo {pages} {3200} (\bibinfo {year}
  {2001})}\BibitemShut {NoStop}%
\bibitem [{\citenamefont {Gleeson}\ and\ \citenamefont
  {Cahalane}(2007)}]{Gleeson-2007}%
  \BibitemOpen
  \bibfield  {author} {\bibinfo {author} {\bibfnamefont {J.~P.}\ \bibnamefont
  {Gleeson}}\ and\ \bibinfo {author} {\bibfnamefont {D.~J.}\ \bibnamefont
  {Cahalane}},\ }\href {\doibase 10.1103/PhysRevE.75.056103} {\bibfield
  {journal} {\bibinfo  {journal} {Phys. Rev. E}\ }\textbf {\bibinfo {volume}
  {75}},\ \bibinfo {pages} {056103} (\bibinfo {year} {2007})}\BibitemShut
  {NoStop}%
\bibitem [{\citenamefont {Jamali}\ and\ \citenamefont
  {Rangwala}(2009)}]{Jamali-2009}%
  \BibitemOpen
  \bibfield  {author} {\bibinfo {author} {\bibfnamefont {S.}~\bibnamefont
  {Jamali}}\ and\ \bibinfo {author} {\bibfnamefont {H.}~\bibnamefont
  {Rangwala}},\ }in\ \href@noop {} {\emph {\bibinfo {booktitle} {Web
  Information Systems and Mining, 2009(WISM 2009)}}}\ (\bibinfo {address}
  {Shanghai, China},\ \bibinfo {year} {2009})\ pp.\ \bibinfo {pages}
  {32--38}\BibitemShut {NoStop}%
\bibitem [{\citenamefont {Kimura}\ \emph {et~al.}(2009)\citenamefont {Kimura},
  \citenamefont {Saito},\ and\ \citenamefont {Motoda}}]{Kimura-2009}%
  \BibitemOpen
  \bibfield  {author} {\bibinfo {author} {\bibfnamefont {M.}~\bibnamefont
  {Kimura}}, \bibinfo {author} {\bibfnamefont {K.}~\bibnamefont {Saito}}, \
  and\ \bibinfo {author} {\bibfnamefont {H.}~\bibnamefont {Motoda}},\ }\href
  {\doibase 10.1145/1514888.1514892} {\bibfield  {journal} {\bibinfo  {journal}
  {ACM Trans. Knowl. Discov. Data}\ }\textbf {\bibinfo {volume} {3}},\ \bibinfo
  {pages} {9} (\bibinfo {year} {2009})}\BibitemShut {NoStop}%
\bibitem [{\citenamefont {Kempe}\ \emph {et~al.}(2003)\citenamefont {Kempe},
  \citenamefont {Kleinberg},\ and\ \citenamefont {Tardos}}]{Kempe-2003}%
  \BibitemOpen
  \bibfield  {author} {\bibinfo {author} {\bibfnamefont {D.}~\bibnamefont
  {Kempe}}, \bibinfo {author} {\bibfnamefont {J.}~\bibnamefont {Kleinberg}}, \
  and\ \bibinfo {author} {\bibfnamefont {{\'E}.}~\bibnamefont {Tardos}},\ }in\
  \href@noop {} {\emph {\bibinfo {booktitle} {Proceedings of the ninth ACM
  SIGKDD International Conference on Knowledge Discovery and Data Mining}}}\
  (\bibinfo {address} {Washington, USA},\ \bibinfo {year} {2003})\ pp.\
  \bibinfo {pages} {137--146}\BibitemShut {NoStop}%
\bibitem [{\citenamefont {Szabo}\ and\ \citenamefont
  {Huberman}(2010)}]{Szabo-2010}%
  \BibitemOpen
  \bibfield  {author} {\bibinfo {author} {\bibfnamefont {G.}~\bibnamefont
  {Szabo}}\ and\ \bibinfo {author} {\bibfnamefont {B.~A.}\ \bibnamefont
  {Huberman}},\ }\href {\doibase 10.1145/1787234.1787254} {\bibfield  {journal}
  {\bibinfo  {journal} {Commun. ACM}\ }\textbf {\bibinfo {volume} {53}},\
  \bibinfo {pages} {80} (\bibinfo {year} {2010})}\BibitemShut {NoStop}%
\bibitem [{\citenamefont {Weng}\ \emph {et~al.}(2013)\citenamefont {Weng},
  \citenamefont {Menczer},\ and\ \citenamefont {Ahn}}]{Weng-2013}%
  \BibitemOpen
  \bibfield  {author} {\bibinfo {author} {\bibfnamefont {L.}~\bibnamefont
  {Weng}}, \bibinfo {author} {\bibfnamefont {F.}~\bibnamefont {Menczer}}, \
  and\ \bibinfo {author} {\bibfnamefont {Y.-Y.}\ \bibnamefont {Ahn}},\ }\href
  {\doibase 10.1038/srep02522} {\bibfield  {journal} {\bibinfo  {journal} {Sci.
  Rep.}\ }\textbf {\bibinfo {volume} {3}},\ \bibinfo {pages} {2522} (\bibinfo
  {year} {2013})}\BibitemShut {NoStop}%
\bibitem [{\citenamefont {Cheng}\ \emph {et~al.}(2014)\citenamefont {Cheng},
  \citenamefont {Adamic}, \citenamefont {Dow}, \citenamefont {Kleinberg},\ and\
  \citenamefont {Leskovec}}]{Cheng-2014}%
  \BibitemOpen
  \bibfield  {author} {\bibinfo {author} {\bibfnamefont {J.}~\bibnamefont
  {Cheng}}, \bibinfo {author} {\bibfnamefont {L.}~\bibnamefont {Adamic}},
  \bibinfo {author} {\bibfnamefont {P.~A.}\ \bibnamefont {Dow}}, \bibinfo
  {author} {\bibfnamefont {J.~M.}\ \bibnamefont {Kleinberg}}, \ and\ \bibinfo
  {author} {\bibfnamefont {J.}~\bibnamefont {Leskovec}},\ }in\ \href@noop {}
  {\emph {\bibinfo {booktitle} {Proceedings of the 23rd International
  Conference on World Wide Web}}}\ (\bibinfo {address} {Seoul, Korea},\
  \bibinfo {year} {2014})\ pp.\ \bibinfo {pages} {925--936}\BibitemShut
  {NoStop}%
\bibitem [{\citenamefont {Yang}\ and\ \citenamefont
  {Leskovec}(Year)}]{Yang-2011}%
  \BibitemOpen
  \bibfield  {author} {\bibinfo {author} {\bibfnamefont {J.}~\bibnamefont
  {Yang}}\ and\ \bibinfo {author} {\bibfnamefont {J.}~\bibnamefont
  {Leskovec}},\ }in\ \href@noop {} {\emph {\bibinfo {booktitle} {Proceedings of
  the fourth ACM International Conference on Web Search and Data Mining}}}\
  (\bibinfo {address} {Hong Kong, China},\ \bibinfo {year} {Year})\ pp.\
  \bibinfo {pages} {177--186}\BibitemShut {NoStop}%
\bibitem [{\citenamefont {Sherchan}\ \emph {et~al.}(2013)\citenamefont
  {Sherchan}, \citenamefont {Nepal},\ and\ \citenamefont
  {Paris}}]{Sherchan-2013}%
  \BibitemOpen
  \bibfield  {author} {\bibinfo {author} {\bibfnamefont {W.}~\bibnamefont
  {Sherchan}}, \bibinfo {author} {\bibfnamefont {S.}~\bibnamefont {Nepal}}, \
  and\ \bibinfo {author} {\bibfnamefont {C.}~\bibnamefont {Paris}},\ }\href
  {\doibase 10.1145/2501654.2501661} {\bibfield  {journal} {\bibinfo  {journal}
  {ACM Comput. Surv.}\ }\textbf {\bibinfo {volume} {45}},\ \bibinfo {pages}
  {47} (\bibinfo {year} {2013})}\BibitemShut {NoStop}%
\bibitem [{\citenamefont {Nowak}\ and\ \citenamefont {May}(1992)}]{NOWAK-1992}%
  \BibitemOpen
  \bibfield  {author} {\bibinfo {author} {\bibfnamefont {M.~A.}\ \bibnamefont
  {Nowak}}\ and\ \bibinfo {author} {\bibfnamefont {R.~M.}\ \bibnamefont
  {May}},\ }\href {\doibase 10.1038/359826a0} {\bibfield  {journal} {\bibinfo
  {journal} {Nature}\ }\textbf {\bibinfo {volume} {359}},\ \bibinfo {pages}
  {826} (\bibinfo {year} {1992})}\BibitemShut {NoStop}%
\bibitem [{\citenamefont {Santos}\ and\ \citenamefont
  {Pacheco}(2005)}]{santos2005scale}%
  \BibitemOpen
  \bibfield  {author} {\bibinfo {author} {\bibfnamefont {F.~C.}\ \bibnamefont
  {Santos}}\ and\ \bibinfo {author} {\bibfnamefont {J.~M.}\ \bibnamefont
  {Pacheco}},\ }\href@noop {} {\bibfield  {journal} {\bibinfo  {journal} {Phys.
  Rev. Lett.}\ }\textbf {\bibinfo {volume} {95}},\ \bibinfo {pages} {098104}
  (\bibinfo {year} {2005})}\BibitemShut {NoStop}%
\bibitem [{\citenamefont {Santos}\ \emph {et~al.}(2006)\citenamefont {Santos},
  \citenamefont {Pacheco},\ and\ \citenamefont
  {Lenaerts}}]{santos2006evolutionary}%
  \BibitemOpen
  \bibfield  {author} {\bibinfo {author} {\bibfnamefont {F.~C.}\ \bibnamefont
  {Santos}}, \bibinfo {author} {\bibfnamefont {J.~M.}\ \bibnamefont {Pacheco}},
  \ and\ \bibinfo {author} {\bibfnamefont {T.}~\bibnamefont {Lenaerts}},\
  }\href@noop {} {\bibfield  {journal} {\bibinfo  {journal} {Proc. Natl. Acad.
  Sci. USA.}\ }\textbf {\bibinfo {volume} {103}},\ \bibinfo {pages} {3490}
  (\bibinfo {year} {2006})}\BibitemShut {NoStop}%
\bibitem [{\citenamefont {Szab{\'o}}\ and\ \citenamefont
  {F\'ath}(2007)}]{GyorgySzab-2007}%
  \BibitemOpen
  \bibfield  {author} {\bibinfo {author} {\bibfnamefont {G.}~\bibnamefont
  {Szab{\'o}}}\ and\ \bibinfo {author} {\bibfnamefont {G.}~\bibnamefont
  {F\'ath}},\ }\href {\doibase 10.1016/j.physrep.2007.04.004} {\bibfield
  {journal} {\bibinfo  {journal} {Phys. Rep.}\ }\textbf {\bibinfo {volume}
  {446}},\ \bibinfo {pages} {97} (\bibinfo {year} {2007})}\BibitemShut
  {NoStop}%
\bibitem [{\citenamefont {Granell}\ \emph {et~al.}(2014)\citenamefont
  {Granell}, \citenamefont {Gomez},\ and\ \citenamefont
  {Arenas}}]{Granell-2014}%
  \BibitemOpen
  \bibfield  {author} {\bibinfo {author} {\bibfnamefont {C.}~\bibnamefont
  {Granell}}, \bibinfo {author} {\bibfnamefont {S.}~\bibnamefont {Gomez}}, \
  and\ \bibinfo {author} {\bibfnamefont {A.}~\bibnamefont {Arenas}},\ }\href
  {\doibase 10.1103/PhysRevE.90.012808} {\bibfield  {journal} {\bibinfo
  {journal} {Phys. Rev. E}\ }\textbf {\bibinfo {volume} {90}},\ \bibinfo
  {pages} {012808} (\bibinfo {year} {2014})}\BibitemShut {NoStop}%
\bibitem [{\citenamefont {Roca}\ \emph {et~al.}(2009)\citenamefont {Roca},
  \citenamefont {Cuesta},\ and\ \citenamefont {Sanchez}}]{Roca-2009}%
  \BibitemOpen
  \bibfield  {author} {\bibinfo {author} {\bibfnamefont {C.~P.}\ \bibnamefont
  {Roca}}, \bibinfo {author} {\bibfnamefont {J.~A.}\ \bibnamefont {Cuesta}}, \
  and\ \bibinfo {author} {\bibfnamefont {A.}~\bibnamefont {Sanchez}},\ }\href
  {\doibase 10.1016/j.plrev.2009.08.001} {\bibfield  {journal} {\bibinfo
  {journal} {Phys. Life Rev.}\ }\textbf {\bibinfo {volume} {6}},\ \bibinfo
  {pages} {208} (\bibinfo {year} {2009})}\BibitemShut {NoStop}%
\bibitem [{\citenamefont {Shakarian}\ \emph {et~al.}(2012)\citenamefont
  {Shakarian}, \citenamefont {Roos},\ and\ \citenamefont
  {Johnson}}]{Shakarian-2012}%
  \BibitemOpen
  \bibfield  {author} {\bibinfo {author} {\bibfnamefont {P.}~\bibnamefont
  {Shakarian}}, \bibinfo {author} {\bibfnamefont {P.}~\bibnamefont {Roos}}, \
  and\ \bibinfo {author} {\bibfnamefont {A.}~\bibnamefont {Johnson}},\ }\href
  {\doibase 10.1016/j.biosystems.2011.09.006} {\bibfield  {journal} {\bibinfo
  {journal} {Biosystems}\ }\textbf {\bibinfo {volume} {107}},\ \bibinfo {pages}
  {66} (\bibinfo {year} {2012})}\BibitemShut {NoStop}%
\bibitem [{\citenamefont {Wang}\ \emph {et~al.}(2012)\citenamefont {Wang},
  \citenamefont {Szolnoki},\ and\ \citenamefont {Perc}}]{wang2012evolution}%
  \BibitemOpen
  \bibfield  {author} {\bibinfo {author} {\bibfnamefont {Z.}~\bibnamefont
  {Wang}}, \bibinfo {author} {\bibfnamefont {A.}~\bibnamefont {Szolnoki}}, \
  and\ \bibinfo {author} {\bibfnamefont {M.}~\bibnamefont {Perc}},\ }\href@noop
  {} {\bibfield  {journal} {\bibinfo  {journal} {EPL-Europhys. Lett.}\ }\textbf
  {\bibinfo {volume} {97}},\ \bibinfo {pages} {48001} (\bibinfo {year}
  {2012})}\BibitemShut {NoStop}%
\bibitem [{\citenamefont {Lieberman}\ \emph {et~al.}(2005)\citenamefont
  {Lieberman}, \citenamefont {Hauert},\ and\ \citenamefont
  {Nowak}}]{Lieberman09}%
  \BibitemOpen
  \bibfield  {author} {\bibinfo {author} {\bibfnamefont {E.}~\bibnamefont
  {Lieberman}}, \bibinfo {author} {\bibfnamefont {C.}~\bibnamefont {Hauert}}, \
  and\ \bibinfo {author} {\bibfnamefont {M.~A.}\ \bibnamefont {Nowak}},\ }\href
  {\doibase 10.1038/nature03204} {\bibfield  {journal} {\bibinfo  {journal}
  {Nature}\ }\textbf {\bibinfo {volume} {433}},\ \bibinfo {pages} {312}
  (\bibinfo {year} {2005})}\BibitemShut {NoStop}%
\bibitem [{\citenamefont {Gleeson}(2008)}]{Gleeson-2008}%
  \BibitemOpen
  \bibfield  {author} {\bibinfo {author} {\bibfnamefont {J.~P.}\ \bibnamefont
  {Gleeson}},\ }\href {\doibase 10.1103/PhysRevE.77.046117} {\bibfield
  {journal} {\bibinfo  {journal} {Phys. Rev. E}\ }\textbf {\bibinfo {volume}
  {77}},\ \bibinfo {pages} {046117} (\bibinfo {year} {2008})}\BibitemShut
  {NoStop}%
\bibitem [{\citenamefont {Melnik}\ \emph {et~al.}(2011)\citenamefont {Melnik},
  \citenamefont {Hackett}, \citenamefont {Porter}, \citenamefont {Mucha},\ and\
  \citenamefont {Gleeson}}]{Melnik-2011}%
  \BibitemOpen
  \bibfield  {author} {\bibinfo {author} {\bibfnamefont {S.}~\bibnamefont
  {Melnik}}, \bibinfo {author} {\bibfnamefont {A.}~\bibnamefont {Hackett}},
  \bibinfo {author} {\bibfnamefont {M.~A.}\ \bibnamefont {Porter}}, \bibinfo
  {author} {\bibfnamefont {P.~J.}\ \bibnamefont {Mucha}}, \ and\ \bibinfo
  {author} {\bibfnamefont {J.~P.}\ \bibnamefont {Gleeson}},\ }\href {\doibase
  10.1103/PhysRevE.83.036112} {\bibfield  {journal} {\bibinfo  {journal} {Phys.
  Rev. E}\ }\textbf {\bibinfo {volume} {83}},\ \bibinfo {pages} {036112}
  (\bibinfo {year} {2011})}\BibitemShut {NoStop}%
\bibitem [{\citenamefont {Yagan}\ and\ \citenamefont
  {Gligor}(2012)}]{Yagan-2012}%
  \BibitemOpen
  \bibfield  {author} {\bibinfo {author} {\bibfnamefont {O.}~\bibnamefont
  {Yagan}}\ and\ \bibinfo {author} {\bibfnamefont {V.}~\bibnamefont {Gligor}},\
  }\href {\doibase 10.1103/PhysRevE.86.036103} {\bibfield  {journal} {\bibinfo
  {journal} {Phys. Rev. E}\ }\textbf {\bibinfo {volume} {86}},\ \bibinfo
  {pages} {036103} (\bibinfo {year} {2012})}\BibitemShut {NoStop}%
\end{thebibliography}

%

\end{document}